# A Bivariate Copula Additive Model for Location, Scale and Shape

Giampiero Marra[*]    Rosalba Radice[†]

Tuesday 5[th] April, 2016

**Abstract**

Rigby & Stasinopoulos (2005) introduced generalized additive models for location, scale and shape (GAMLSS) where the response distribution is not restricted to belong to the exponential family and its parameters can be specified as functions of additive predictors that allows for several types of covariate effects (e.g., linear, non-linear, random and spatial effects). In many empirical situations, however, modeling simultaneously two or more responses conditional on some covariates can be of considerable relevance. In this article, we extend the scope of GAMLSS by introducing a bivariate copula additive model with continuous margins for location, scale and shape. The framework permits the copula dependence and marginal distribution parameters to be estimated simultaneously and, like in GAMLSS, each parameter to be modeled using an additive predictor. Parameter estimation is achieved within a penalized likelihood framework using a trust region algorithm with integrated automatic multiple smoothing parameter selection. The proposed approach allows for straightforward inclusion of potentially any parametric continuous marginal distribution and copula function. The models can be easily used via the `copulaReg()` function in the R package `SemiParBIVProbit`. The usefulness of the proposal is illustrated on two case studies (which use electricity price and demand data, and birth records) and on simulated data.

**Key Words**: additive predictor, continuous marginal distribution, copula, simultaneous parameter estimation.

---

[*]Department of Statistical Science, University College London, Gower Street, London WC1E 6BT, UK, `giampiero.marra@ucl.ac.uk`.

[†]Department of Economics, Mathematics and Statistics, Birkbeck, University of London, Malet Street, London WC1E 7HX, UK, `r.radice@bbk.ac.uk`.

# 1  Introduction

Regression models typically involve a response variable and a set of covariates. However, modeling simultaneously two or more responses conditional on some covariates can be of considerable empirical relevance. Some examples can be drawn from health economics (e.g., modeling self-selection and dependence between health insurance and health care demand among married couples), engineering and econometrics (e.g., building time-series models for electricity price and demand), biostatistics (e.g., modeling adverse birth outcomes), actuarial science (e.g., studying the interdependence between mortality and losses) and finance (e.g., modeling jointly the prices of different assets); see Trivedi & Zimmer (2006) for details and more examples. The copula approach offers a convenient and computationally tractable framework to model multivariate responses in a regression context and it has been the subject of many methodological developments over the last few years (e.g., Cherubini et al., 2004; Kolev & Paiva, 2009; Nelsen, 2006; Radice et al., 2015).

Rigby & Stasinopoulos (2005) extended the class of generalized additive models (Hastie & Tibshirani, 1990; Wood, 2006) by introducing generalized additive models for location, scale and shape (GAMLSS). Here, the response distribution is not restricted to belong to the exponential family and its parameters can be modeled using flexible functions of explanatory variables. In this article, we extend the scope of GAMLSS by introducing a bivariate copula additive model with continuous margins for location, scale and shape. The framework permits the copula dependence and marginal distribution parameters to be estimated simultaneously and, like in GAMLSS, each parameter to be specified as a function of an additive predictor incorporating several types of covariate effects (e.g., linear, non-linear, random and spatial effects). The method allows for the use of potentially any parametric continuous marginal response distribution (eleven distributions have been implemented for this work), several dependence structures between the margins as implied by copulae, and whenever appropriate rotated versions of them (seven copulae have been considered here), as many additive predictors as the number of parameters of the marginal distributions and copula. Our proposal can also be regarded as an extension of the copula models introduced by Radice et al. (2015) as well as those implemented in the VGAM R package (Yee, 2016) and as a frequentist counterpart of the approach by Klein & Kneib (2015). Furthermore, to the best of our knowledge, other existing bivariate copula regression approaches and software implementations cover only parts of the flexibility of the proposed approach (see, e.g., Acar et al., 2013;

Craiu & Sabeti, 2012; Gijbels et al., 2011; Kramer et al., 2012; Kraemer & Silvestrini, 2015; Yan, 2007, and references therein).

The reader is cautioned that the methodology developed in this article is most useful when the main interest is in relating the parameters of a bivariate copula distribution to covariate effects. Otherwise, semi/non-parametric extensions where, for instance, the margins and/or copula function are estimated using kernels, wavelets or orthogonal polynomials may be considered instead (e.g., Kauermann et al., 2013; Lambert, 2007; Segers et al., 2014; Shen et al., 2008). Such techniques are in principle more flexible in determining the shape of the underlying bivariate distribution. In practice, however, they are very limited with regard to the inclusion of flexible covariate effects, may require the imposition of identifying restrictions on the functions approximating the underlying distribution and may require large sample sizes to produce reliable results.

The remainder of the paper is organized as follows. Section 2 introduces the proposed class of bivariate copula additive models by describing its main building blocks. Section 3 provides some estimation details and discusses the modularity of the implementation. Section 4 gives further details on the modeling framework, whereas Section 5 provides some guidelines for model building. Section 6 illustrates the usefulness of the approach on two empirical case studies (which use time series and cross sectional data) and on simulated data. Section 7 discusses some potential directions for future research.

The models discussed in this paper can be easily used via the `copulaReg()` function in the R package `SemiParBIVProbit` (Marra & Radice, 2016), and the reader can reproduce the analyzes in this paper using the R scripts in the supplementary material.

## 2 Copula additive model for location, scale and shape

Let $\mathcal{F}(y_1, y_2 | \mathbf{z}_1, \mathbf{z}_2)$ denote the joint cumulative distribution function (cdf) of two continuous random variables, $Y_1$ and $Y_2$, conditional on the sets of covariates $\mathbf{z}_1$ and $\mathbf{z}_2$ (note that $\mathbf{z}_1$ and $\mathbf{z}_2$ need not be different sets of covariates). It is possible to state that

$$\mathcal{F}(y_1, y_2 | \mathbf{z}_1, \mathbf{z}_2) = \mathcal{C}\left(\mathcal{F}_1(y_1 | \mathbf{z}_1), \mathcal{F}_2(y_2 | \mathbf{z}_2); \theta\right), \qquad (1)$$

where $\mathcal{F}_1(y_1|\mathbf{z}_1)$ and $\mathcal{F}_2(y_2|\mathbf{z}_2)$ are marginal cdfs of $Y_1$ and $Y_2$ taking values in $(0,1)$ which are specified to be conditioned on $\mathbf{z}_1$ and $\mathbf{z}_2$, $\mathcal{C}(\cdot,\cdot)$ is a uniquely defined two-place copula function which does not depend on the marginal cdfs and $\theta$ is an association copula parameter measuring the dependence between the two marginals (Kolev & Paiva, 2009; Sklar, 1959, 1973; Zimmer & Trivedi, 2006). A substantial advantage of the copula approach is that a joint cdf can be conveniently expressed in terms of (arbitrary) univariate marginal cdfs and a function $\mathcal{C}$ that binds them together. The copulae implemented in `SemiParBIVProbit` are reported in Table 1. Rotated versions of the Clayton, Gumbel and Joe copulae can also be obtained. Specifically, rotation by 180 degrees leads to the survival copula ($\mathcal{C}_{180}$), while rotation by 90 ($\mathcal{C}_{90}$) and 270 degrees ($\mathcal{C}_{270}$) allows for negative dependence which is not possible with the non-rotated and survival versions. Following Brechmann & Schepsmeier (2013), these are defined as $\mathcal{C}_{90}(u,v) = v - \mathcal{C}(1-u,v)$, $\mathcal{C}_{180}(u,v) = u + v - 1 + \mathcal{C}(1-u, 1-v)$ and $\mathcal{C}_{270}(u,v) = u - \mathcal{C}(u, 1-v)$, where $\theta$ has been dropped for notational convenience, $u = \mathcal{F}_1(y_1|\mathbf{z}_1)$ and $v = \mathcal{F}_2(y_2|\mathbf{z}_2)$. As shown in Table 1, for each copula there exists a relation between $\theta$ and the Kendall's $\tau$ coefficient, which is a convenient measure of association that lies in the customary range $[-1,1]$. More details on copulae can be found in Nelsen (2006) and Trivedi & Zimmer (2006).

The marginal distributions of $Y_1$ and $Y_2$ are specified through parametric cdfs and densities which can be precisely denoted as $\mathcal{F}_m(y_m|\mu_m, \sigma_m, \nu_m)$ and $f_m(y_m|\mu_m, \sigma_m, \nu_m)$, for $m = 1, 2$, where $\mu_m$, $\sigma_m$ and $\nu_m$ are marginal distribution parameters (which usually represent location, scale and shape) which can be related to predictors (Rigby & Stasinopoulos, 2005). The number of coefficients that characterize $\mathcal{F}_m$ and $f_m$ depends on the chosen distribution and we have considered the two and three parameter distributions described in Table 2.

Finally, as suggested above, all (marginal distribution and dependence) parameters are related to additive predictors $\eta$'s (defined in generic terms in the next section) via known monotonic link functions which ensure that the restrictions on the parameter spaces are maintained (see Table 1 and the caption of Table 2 for the transformations employed). For example, if $\sigma_1$ and $\sigma_2$ can only take positive values and we wish to model them as functions of covariates (which can be useful to capture non-homogeneity in the parameters) then we can specify $\log(\sigma_1) = \eta_{\sigma_1}$ and $\log(\sigma_2) = \eta_{\sigma_2}$. As for the copula parameter, we can use, for example, $\log(\theta - 1) = \eta_\theta$ in the Gumbel case which allows for the strength of the (upper tail) dependence between the marginals

| Copula | $\mathcal{C}(u, v; \theta)$ | Range of $\theta$ | Transformation of $\theta$ | Kendall's $\tau$ |
|---|---|---|---|---|
| AMH (`"AMH"`) | $\frac{uv}{1-\theta(1-u)(1-v)}$ | $\theta \in [-1,1]$ | $\tanh^{-1}(\theta)$ | $1 - \frac{2}{3\theta^2}\{\theta + (1-\theta)^2 \log(1-\theta)\}$ |
| Clayton (`"C0"`) | $(u^{-\theta} + v^{-\theta} - 1)^{-1/\theta}$ | $\theta \in (0, \infty)$ | $\log(\theta - \epsilon)$ | $\frac{\theta}{\theta+2}$ |
| FGM (`"FGM"`) | $uv\{1 + \theta(1-u)(1-v)\}$ | $\theta \in [-1,1]$ | $\tanh^{-1}(\theta)$ | $\frac{2}{9}\theta$ |
| Frank (`"F"`) | $-\theta^{-1}\log\{1 + (e^{-\theta u} - 1)(e^{-\theta v} - 1)/(e^{-\theta} - 1)\}$ | $\theta \in \mathbb{R}\setminus\{0\}$ | – | $1 - \frac{4}{\theta}[1 - D_1(\theta)]$ |
| Gaussian (`"N"`) | $\Phi_2(\Phi^{-1}(u), \Phi^{-1}(v); \theta)$ | $\theta \in [-1,1]$ | $\tanh^{-1}(\theta)$ | $\frac{2}{\pi}\arcsin(\theta)$ |
| Gumbel (`"G0"`) | $\exp[-\{(-\log u)^\theta + (-\log v)^\theta\}^{1/\theta}]$ | $\theta \in [1, \infty)$ | $\log(\theta - 1)$ | $1 - \frac{1}{\theta}$ |
| Joe (`"J0"`) | $1 - \{(1-u)^\theta + (1-v)^\theta - (1-u)^\theta(1-v)^\theta\}^{1/\theta}$ | $\theta \in (1, \infty)$ | $\log(\theta - 1 - \epsilon)$ | $1 + \frac{4}{\theta^2}D_2(\theta)$ |

Table 1: Definition of copulae implemented in `SemiParBIVProbit`, with corresponding parameter range of association parameter $\theta$, transformation of $\theta$ and relation between Kendall's $\tau$ and $\theta$. $\Phi_2(\cdot, \cdot; \theta)$ denotes the cdf of a standard bivariate normal distribution with correlation coefficient $\theta$, and $\Phi(\cdot)$ the cdf of a univariate standard normal distribution. $D_1(\theta) = \frac{1}{\theta}\int_0^\theta \frac{t}{\exp(t)-1}dt$ is the Debye function and $D_2(\theta) = \int_0^1 t \log(t)(1-t)^{\frac{2(1-\theta)}{\theta}}dt$. Quantity $\epsilon$ is set to the machine smallest positive floating-point number multiplied by $10^6$ and is used to ensure that the restrictions on the space of $\theta$ are maintained. Argument `BivD` of `copulaReg()` in `SemiParBIVProbit` allows the user to employ the desired copula function and can be set to any of the values within brackets next to the copula names in the first column; for example, `BivD = "J0"`. For Clayton, Gumbel and Joe, the number after the capital letter indicates the degree of rotation required: the possible values are 0, 90, 180 and 270. Recall, for instance, that the Gaussian and Frank copulae allow for equal degrees of positive and negative dependence, with Frank exhibiting a slightly stronger dependence in the middle of the distribution. Clayton is asymmetric with a strong lower tail dependence but a weaker upper tail dependence, and vice-versa for the Gumbel and Joe copulae.

to vary across observations.

In the above, observation index $i$ has been suppressed to avoid clutter in the notation. However, it should be clear from the context of the paper that the focus is on modeling independent bivariate realizations $(y_{i1}, y_{i2})^\mathsf{T}$ as functions of $\mathbf{z}_{1i}$ and $\mathbf{z}_{2i}$, where $i = 1, \ldots, n$ and $n$ is the sample size.

## 2.1 Predictor specification

Let us define a generic predictor $\eta_i$ as a function of an intercept and smooth functions of sub-vectors of a generic covariate vector $\mathbf{z}_i$. That is,

$$\eta_i = \beta_0 + \sum_{k=1}^{K} s_k(\mathbf{z}_{ki}), \ i = 1, \ldots, n, \quad (2)$$

where $\beta_0 \in \mathbb{R}$ is an overall intercept, $\mathbf{z}_{ki}$ denotes the $k^{th}$ sub-vector of the complete covariate vector $\mathbf{z}_i$ (containing, e.g., binary, categorical, continuous, and spatial variables) and the $K$ functions $s_k(\mathbf{z}_{ki})$ represent generic effects which are chosen according to the type of covariate(s) considered. Each $s_k(\mathbf{z}_{ki})$ can be approximated as a linear combination of $J_k$ basis functions $b_{kj_k}(\mathbf{z}_{ki})$ and

| | $\mathcal{F}_m(y_m|\mu_m,\sigma_m,\nu_m)$ | $f_m(y_m|\mu_m,\sigma_m,\nu_m)$ | $\mathbb{E}(Y_m)$ | $\mathbb{V}(Y_m)$ | Support of $y_m$ / Parameter ranges |
|---|---|---|---|---|---|
| beta ("BE") | $I(y;\alpha_1,\alpha_2)$<br>$\alpha_1 = \frac{\mu(1-\sigma^2)}{\sigma^2}$<br>$\alpha_2 = \frac{(1-\mu)(1-\sigma^2)}{\sigma^2}$ | $\frac{y^{\alpha_1-1}(1-y)^{\alpha_2-1}}{B(\alpha_1,\alpha_2)}$ | $\mu$ | $\sigma^2\mu(1-\mu)$ | $0 < y < 1$<br>$0 < \mu < 1, 0 < \sigma < 1$ |
| Dagum ("DAGUM") | $\left\{1 + \left(\frac{y}{\mu}\right)^{-\sigma}\right\}^{-\nu}$ | $\frac{\sigma\nu}{y}\left[\frac{\left(\frac{y}{\mu}\right)^{\sigma\nu}}{\left\{\left(\frac{y}{\mu}\right)^\sigma+1\right\}^{\nu+1}}\right]$ | $-\frac{\mu}{\sigma}\frac{\Gamma(-\frac{1}{\sigma})\Gamma(\frac{1}{\sigma}+\nu)}{\Gamma(\nu)}$<br>if $\sigma > 1$ | $-\left(\frac{\mu}{\sigma}\right)^2\left[2\sigma\frac{\Gamma(-\frac{2}{\sigma})\Gamma(\frac{2}{\sigma}+\nu)}{\Gamma(\nu)}\right.$<br>$\left.+\left\{\frac{\Gamma(-\frac{1}{\sigma})\Gamma(\frac{1}{\sigma}+\nu)}{\Gamma(\nu)}\right\}^2\right]$<br>if $\sigma > 2$ | $y > 0$<br>$\mu > 0, \sigma > 0, \nu > 0$ |
| gamma ("GA") | $\frac{1}{\Gamma(\frac{1}{\sigma^2})}\gamma\left(\frac{1}{\sigma^2}, \frac{y}{\mu\sigma^2}\right)$ | $\frac{1}{(\mu\sigma^2)^{\frac{1}{\sigma^2}}}\frac{y^{\frac{1}{\sigma^2}-1}\exp\left(-\frac{y}{\mu\sigma^2}\right)}{\Gamma(\frac{1}{\sigma^2})}$ | $\mu$ | $\mu^2\sigma^2$ | $y > 0$<br>$\mu > 0, \sigma > 0$ |
| Gumbel ("GU") | $1 - \exp\left\{-\exp\left(\frac{y-\mu}{\sigma}\right)\right\}$ | $\frac{1}{\sigma}\exp\left\{\left(\frac{y-\mu}{\sigma}\right) - \exp\left(\frac{y-\mu}{\sigma}\right)\right\}$ | $\mu - 0.57722\sigma$ | $\frac{\pi^2\sigma^2}{6}$ | $-\infty < y < \infty$<br>$-\infty < \mu < \infty, \sigma > 0$ |
| inverse Gaussian ("iG") | $\Phi\left\{\frac{1}{\sqrt{y\sigma^2}}\left(\frac{y}{\mu}-1\right)\right\}+$<br>$\exp\left(\frac{2}{\mu\sigma^2}\right)$<br>$\Phi\left\{-\frac{1}{\sqrt{y\sigma^2}}\left(\frac{y}{\mu}+1\right)\right\}$ | $\frac{1}{\sqrt{2\pi\sigma^2 y^3}}\exp\left\{-\frac{1}{2\mu^2\sigma^2 y}(y-\mu)^2\right\}$ | $\mu$ | $\mu^3\sigma^2$ | $y > 0$<br>$\mu > 0, \sigma > 0$ |
| log-normal ("LN") | $\frac{1}{2} + \frac{1}{2}\mathrm{erf}\left\{\frac{\log(y)-\mu}{\sigma\sqrt{2}}\right\}$ | $\frac{1}{y\sigma\sqrt{2\pi}}\exp\left[-\frac{\{\log(y)-\mu\}^2}{2\sigma^2}\right]$ | $\sqrt{\exp(\sigma^2)}\exp(\mu)$ | $\exp(\sigma^2)\{\exp(\sigma^2)-1\}\exp(2\mu)$ | $y > 0$<br>$\mu > 0, \sigma > 0$ |
| logistic ("LO") | $\frac{1}{1+\exp(-\frac{y-\mu}{\sigma})}$ | $\frac{1}{\sigma}\left\{\exp\left(-\frac{y-\mu}{\sigma}\right)\right\}\left\{1+\exp\left(-\frac{y-\mu}{\sigma}\right)\right\}^{-2}$ | $\mu$ | $\frac{\pi^2\sigma^2}{3}$ | $-\infty < y < \infty$<br>$-\infty < \mu < \infty, \sigma > 0$ |
| normal ("N") | $\frac{1}{2}\left\{1 + \mathrm{erf}\left(\frac{y-\mu}{\sigma\sqrt{2}}\right)\right\}$ | $\frac{1}{\sigma\sqrt{2\pi}}\exp\left\{-\frac{(y-\mu)^2}{2\sigma^2}\right\}$ | $\mu$ | $\sigma^2$ | $-\infty < y < \infty$<br>$-\infty < \mu < \infty, \sigma > 0$ |
| reverse Gumbel ("rGU") | $\exp\left\{-\exp\left(-\frac{y-\mu}{\sigma}\right)\right\}$ | $\frac{1}{\sigma}\exp\left\{\left(-\frac{y-\mu}{\sigma}\right) - \exp\left(-\frac{y-\mu}{\sigma}\right)\right\}$ | $\mu + 0.57722\sigma$ | $\frac{\pi^2\sigma^2}{6}$ | $-\infty < y < \infty$<br>$-\infty < \mu < \infty, \sigma > 0$ |
| Singh-Maddala ("SM") | $1 - \left\{1 + \left(\frac{y}{\mu}\right)^\sigma\right\}^{-\nu}$ | $\frac{\sigma\nu y^{\sigma-1}}{\mu^\sigma\{1+(\frac{y}{\mu})^\sigma\}^{\nu+1}}$ | $\mu\frac{\Gamma(1+\frac{1}{\sigma})\Gamma(-\frac{1}{\sigma}+\nu)}{\Gamma(\nu)}$<br>if $\sigma\nu > 1$ | $\mu^2\left\{\Gamma\left(1+\frac{2}{\sigma}\right)\Gamma(\nu)\Gamma\left(-\frac{2}{\sigma}+\nu\right)\right.$<br>$\left.-\Gamma\left(1+\frac{1}{\sigma}\right)^2\Gamma\left(-\frac{1}{\sigma}+\nu\right)^2\right\}$<br>if $\sigma\nu > 2$ | $y > 0$<br>$\mu > 0, \sigma > 0, \nu > 0$ |
| Weibull ("WEI") | $1 - \exp\left\{-\left(\frac{y}{\mu}\right)^\sigma\right\}$ | $\frac{\sigma}{\mu}\left(\frac{y}{\mu}\right)^{\sigma-1}\exp\left\{-\left(\frac{y}{\mu}\right)^\sigma\right\}$ | $\mu\Gamma\left(\frac{1}{\sigma}+1\right)$ | $\mu^2\left[\Gamma\left(\frac{2}{\sigma}+1\right) - \left\{\Gamma\left(\frac{1}{\sigma}+1\right)\right\}^2\right]$ | $y > 0$<br>$\mu > 0, \sigma > 0$ |

Table 2: Definition and some properties of the distributions implemented in SemiParBIVProbit. Following the parametrization and convention adopted by Rigby & Stasinopoulos (2005), these are defined in terms of $\mu$, $\sigma$ and $\nu$ which in most cases represent location, scale and shape. Subscript $m$ can take values 1 and 2; to avoid clutter in the notation we have suppressed $m$ in the main body of the table. The means and variances of DAGUM and SM are indeterminate for certain values of $\sigma$ and $\nu$. If a parameter can only take positive values then transformation $\log(\eta - \epsilon)$ is employed, where $\eta$ is a linear predictor defined in Section 2.1 and $\epsilon$ is defined and its use explained in the caption of Table 1. If a parameter can take values in $(0,1)$ then the inverse of the cumulative distribution function of a standardized logistic is used. $I(\cdot;\cdot,\cdot)$ is the regularized beta function, $B(\cdot,\cdot)$ is the beta function, $\Gamma(\cdot)$ is the gamma function, $\gamma(\cdot,\cdot)$ is the lower incomplete gamma function, $\Phi(\cdot)$ is the cdf of a univariate standard normal distribution and erf$(\cdot)$ is the error function. Argument margins of copulaReg() in SemiParBIVProbit allows the user to employ the desired marginal distributions and can be set to any of the values within brackets next to the names in the first column; for example, margins = c("WEI", "DAGUM"). Note that in many cases the parameters of the distributions determine $\mathbb{E}(y_m)$ and $\mathbb{V}(y_m)$ through functions of them. Also, if $Y^*$ is log-normally distributed then $Y = \log(Y^*)$ has a normal distribution; likewise, if $Y$ has a normal distribution then $Y^* = \exp(Y)$ has a log-normal distribution. If $Y \sim \text{rGU}(\mu,\sigma)$ and $Y^* = -Y$ then $Y^* \sim \text{GU}(-\mu,\sigma)$.

regression coefficients $\beta_{kj_k} \in \mathbb{R}$, i.e.

$$\sum_{j_k=1}^{J_k} \beta_{kj_k} b_{kj_k}(\mathbf{z}_{ki}). \tag{3}$$

Equation (3) implies that the vector of evaluations $\{s_k(\mathbf{z}_{k1}), \ldots, s_k(\mathbf{z}_{kn})\}^\mathsf{T}$ can be written as $\mathbf{Z}_k \boldsymbol{\beta}_k$ with $\boldsymbol{\beta}_k = (\beta_{k1}, \ldots, \beta_{kJ_k})^\mathsf{T}$ and design matrix $Z_k[i, j_k] = b_{kj_k}(\mathbf{z}_{ki})$. This allows the predictor in equation (2) to be written as

$$\boldsymbol{\eta} = \beta_0 \mathbf{1}_n + \mathbf{Z}_1 \boldsymbol{\beta}_1 + \ldots + \mathbf{Z}_K \boldsymbol{\beta}_K, \tag{4}$$

where $\mathbf{1}_n$ is an $n$-dimensional vector made up of ones. Equation (4) can also be written in a more compact way as $\boldsymbol{\eta} = \mathbf{Z}\boldsymbol{\beta}$, where $\mathbf{Z} = (\mathbf{1}_n, \mathbf{Z}_1, \ldots, \mathbf{Z}_K)$ and $\boldsymbol{\beta} = (\beta_0, \boldsymbol{\beta}_1^\mathsf{T}, \ldots, \boldsymbol{\beta}_K^\mathsf{T})^\mathsf{T}$. The smooth functions may represent linear, non-linear, random and spatial effects, to name but a few. Moreover, each $\boldsymbol{\beta}_k$ has an associated quadratic penalty $\lambda_k \boldsymbol{\beta}_k^\mathsf{T} \mathbf{D}_k \boldsymbol{\beta}_k$ whose role is to enforce specific properties on the $k^{th}$ function, such as smoothness. Smoothing parameter $\lambda_k \in [0, \infty)$ controls the trade-off between fit and smoothness, and plays a crucial role in determining the shape of $\hat{s}_k(\mathbf{z}_{ki})$. For instance, let us assume that the $k^{th}$ function models the effect of a continuous variable. A value of $\lambda_k = 0$ (i.e., no penalization is applied to $\boldsymbol{\beta}_k$ during fitting) will result in an un-penalized regression spline estimate with the likely consequence of over-fitting, while $\lambda_k \to \infty$ (i.e., the penalty has a large influence on $\boldsymbol{\beta}_k$ during fitting) will lead to a straight line estimate. The overall penalty can be defined as $\boldsymbol{\beta}^\mathsf{T} \mathbf{D}_{\boldsymbol{\lambda}} \boldsymbol{\beta}$, where $\mathbf{D}_{\boldsymbol{\lambda}} = \text{diag}(0, \lambda_1 \mathbf{D}_1, \ldots, \lambda_K \mathbf{D}_K)$. The smooth functions are typically subject to centering (identifiability) constraints and we follow the parsimonious approach detailed in Wood (2006) to deal with this issue. In the following paragraphs, we discuss some smooth function specifications.

**Linear and random effects**  For parametric, linear effects, equation (3) becomes $\mathbf{z}_{ki}^\mathsf{T} \boldsymbol{\beta}_k$, and the design matrix is obtained by stacking all covariate vectors $\mathbf{z}_{ki}$ into $\mathbf{Z}_k$. No penalty is typically assigned to linear effects ($\mathbf{D}_k = \mathbf{0}$). This would be the case for binary and categorical variables. However, sometimes it is desirable to penalize parametric linear effects. For instance, the coefficients of some factor variables in the model may be weakly or not identified by the data. In this case, a ridge penalty could be employed to make the model parameters estimable (here $\mathbf{D}_k = \mathbf{I}$ where **I** is an identity matrix). This is equivalent to the assumption that the coefficients are *i.i.d.* normal random effects with unknown variance (e.g., Ruppert et al., 2003; Wood, 2006). An example of specification of an equation containing two factor variables, one of which requires the use of a ridge penalty is

```
y ~ x1 + s(x2, bs = "re")
```

where `y` is a response, and `x1` and `x2` are factor variables. Argument `bs` specifies the type of spline basis employed which in this case is `re` (random effect).

**Non-linear effects** For continuous variables the smooth functions are represented using the regression spline approach popularized by Eilers & Marx (1996). Specifically, for each continuous variable $z_{ki}$, $s_k(z_{ki})$ is approximated by $\sum_{j_k=1}^{J_k} \beta_{kj_k} b_{kj_k}(z_{ki})$, where the $b_{kj_k}(z_{ki})$ are known spline basis functions. The design matrix $\mathbf{Z}_k$ comprises the basis function evaluations for each $i$, and hence describe $J_k$ curves which have potentially varying degrees of complexity. We employ low rank thin plate regression splines (Wood, 2003) which are numerically stable and have convenient mathematical properties, although other spline definitions and corresponding penalties are supported in our implementation. Note that for one-dimensional smooth functions, the choice of spline definition does not play an important role in determining the shape of $\hat{s}_k(z_k)$ (e.g., Ruppert et al., 2003). To enforce smoothness, a conventional integrated square second derivative spline penalty is typically employed (this is also the default option in the software). That is, $\mathbf{D}_k = \int \mathbf{d}_k(z_k) \mathbf{d}_k(z_k)^\mathsf{T} dz_k$, where the $j_k^{th}$ element of $\mathbf{d}_k(z_k)$ is given by $\partial^2 b_{kj_k}(z_k)/\partial z_k^2$ and integration is over the range of $z_k$. The formulae used to compute the basis functions and penalties for many spline definitions are provided in Ruppert et al. (2003) and Wood (2006). For their theoretical properties see, for instance, Kauermann et al. (2009) and Yoshida & Naito (2014). This specification allows us to avoid arbitrary modeling decisions, such as choosing the appropriate degree of a polynomial or specifying cut-points, which could induce misspecification bias. The example of the previous paragraph can be extended to include `s(x3, bs = "tp", k = 10)`, where `x3` is a continuous covariate and `bs` is set to `tp` (penalized low rank thin plate spline, the default) with `k = 10` number of basis functions. Argument `bs` can also be set to, for example, `cr` (penalized cubic regression spline) and `ps` (P-spline).

**Spatial effects**   When the geographic area (or country) of interest is split up into discrete contiguous geographic units (or regions) and such information is available, a Markov random field approach can be employed to exploit the information contained in neighboring observations which are located in the same country. In this case, equation (3) becomes $\mathbf{z}_{ki}^\mathsf{T}\boldsymbol{\beta}_k$ where $\boldsymbol{\beta}_k = (\beta_{k1}, \ldots, \beta_{kR})^\mathsf{T}$ represents the vector of spatial effects, $R$ denotes the total number of regions and $\mathbf{z}_{ki}$ is made up of a set of area labels. The design matrix linking an observation $i$ to the corresponding spatial effect is therefore defined as

$$\mathbf{Z}_k[i,r] = \begin{cases} 1 & \text{if the observation belongs to region } r \\ 0 & \text{otherwise} \end{cases},$$

where $r = 1, \ldots, R$. The smoothing penalty is based on the neighborhood structure of the geographic units, so that spatially adjacent regions share similar effects. That is,

$$\mathbf{D}_k[r,q] = \begin{cases} -1 & \text{if } r \neq q \wedge r \sim q \\ 0 & \text{if } r \neq q \wedge r \nsim q \\ N_r & \text{if } r = q \end{cases},$$

where $r \sim q$ indicates whether two regions $r$ and $q$ are adjacent neighbors, and $N_r$ is the total number of neighbors for region $r$. In a stochastic interpretation, this penalty is equivalent to the assumption that $\boldsymbol{\beta}_k$ follows a Gaussian Markov random field (e.g., Rue & Held, 2005). The above example can be further extended to include `s(x4, bs = "mrf")` where `x4` is a factor variable and `mrf` stands for Markov random field.

Several other specifications can be employed. These include varying coefficient smooths obtained by multiplying one or more smooth components by some covariate(s), and smooth functions of two or more continuous covariates (e.g., Wood, 2006). The smoothers utilized here are obtained from the R package `mgcv` package whose documentation can be consulted for more details (Wood, 2016).

# 3 Log-likelihood and some estimation details

For notational convenience, let us suppress for a moment the conditioning on covariates and parameters. If $\mathcal{F}_1$ and $\mathcal{F}_2$ are continuous with densities $f_1$ and $f_2$ then joint density $f$, resulting from equation (1), is given by

$$f(y_{1i}, y_{2i}) = \frac{\partial^2 \mathcal{F}(y_{1i}, y_{2i})}{\partial y_{1i} \partial y_{2i}} = \frac{\partial^2 \mathcal{C}\left(\mathcal{F}_1(y_{1i}), \mathcal{F}_2(y_{2i})\right)}{\partial \mathcal{F}_1(y_{1i}) \partial \mathcal{F}_2(y_{2i})} \times \frac{\partial \mathcal{F}_1(y_{1i})}{\partial y_{1i}} \times \frac{\partial \mathcal{F}_2(y_{2i})}{\partial y_{2i}},$$

which, for the overall parameter vector $\boldsymbol{\delta}$ (defined in the next paragraph), can be re-written as

$$f(y_{1i}, y_{2i}|\boldsymbol{\delta}) = c\left(\mathcal{F}_1(y_{1i}|\mu_{1i}, \sigma_{1i}, \nu_{1i}), \mathcal{F}_2(y_{2i}|\mu_{2i}, \sigma_{2i}, \nu_{2i}); \theta_i\right) f_1(y_{1i}|\mu_{1i}, \sigma_{1i}, \nu_{1i}) f_2(y_{2i}|\mu_{2i}, \sigma_{2i}, \nu_{2i}),$$

where $c(\cdot, \cdot; \cdot)$ is the copula density. Therefore, the log-likelihood function is

$$\ell(\boldsymbol{\delta}) = \sum_{i=1}^{n} \log\left\{c\left(\mathcal{F}_1(y_{1i}|\mu_{1i}, \sigma_{1i}, \nu_{1i}), \mathcal{F}_2(y_{2i}|\mu_{2i}, \sigma_{2i}, \nu_{2i}); \theta_i\right)\right\} + \sum_{i=1}^{n} \sum_{m=1}^{2} \log\left\{f_m(y_{mi}|\mu_{mi}, \sigma_{mi}, \nu_{mi})\right\},$$

where parameter vector $\boldsymbol{\delta}$ is defined as $(\boldsymbol{\beta}_{\mu_1}^\mathsf{T}, \boldsymbol{\beta}_{\mu_2}^\mathsf{T}, \boldsymbol{\beta}_{\sigma_1}^\mathsf{T}, \boldsymbol{\beta}_{\sigma_2}^\mathsf{T}, \boldsymbol{\beta}_{\nu_1}^\mathsf{T}, \boldsymbol{\beta}_{\nu_2}^\mathsf{T}, \boldsymbol{\beta}_{\theta}^\mathsf{T})^\mathsf{T}$ when three parameter distributions are employed for both margins; the parameter vectors that make up $\boldsymbol{\delta}$ relate to $\eta_{\mu_1 i}$, $\eta_{\mu_2 i}$, $\eta_{\sigma_1 i}$, $\eta_{\sigma_2 i}$, $\eta_{\nu_1 i}$, $\eta_{\nu_2 i}$ and $\eta_{\theta_i}$. Because of the flexible predictors' structures employed here, the use of a classic (unpenalized) optimization algorithm is likely to result in smooth function estimates which may not reflect the true underlying trends in the data (e.g., Ruppert et al., 2003; Wood, 2006). Therefore, we maximize

$$\ell_p(\boldsymbol{\delta}) = \ell(\boldsymbol{\delta}) - \frac{1}{2} \boldsymbol{\delta}^\mathsf{T} \mathbf{S}_{\boldsymbol{\lambda}} \boldsymbol{\delta}, \tag{5}$$

where $\mathbf{S}_{\boldsymbol{\lambda}} = \text{diag}(\boldsymbol{\lambda}_{\mu_1} \mathbf{D}_{\mu_1}, \boldsymbol{\lambda}_{\mu_2} \mathbf{D}_{\mu_2}, \boldsymbol{\lambda}_{\sigma_1} \mathbf{D}_{\sigma_1}, \boldsymbol{\lambda}_{\sigma_2} \mathbf{D}_{\sigma_2}, \boldsymbol{\lambda}_{\nu_1} \mathbf{D}_{\nu_1}, \boldsymbol{\lambda}_{\nu_2} \mathbf{D}_{\nu_2}, \boldsymbol{\lambda}_{\theta} \mathbf{D}_{\theta})$ with each generic $\boldsymbol{\lambda}$ defined as $(\lambda_1, \ldots, \lambda_K)^\mathsf{T}$. If two and three parameter distributions or two parameter distributions are employed then $\boldsymbol{\delta}$ and $\mathbf{S}_{\boldsymbol{\lambda}}$ have to be re-defined in the obvious way.

To maximize (5), we have extended the efficient and stable trust region algorithm with integrated automatic multiple smoothing parameter selection introduced by Radice et al. (2015) to incorporate potentially any parametric continuous marginal distribution and one-parameter copula function, and to link all parameters of the model to additive predictors; a sketch of the algorithm is given in supplementary material Section 1 (SM-1). Starting values for the parameters of the marginals are obtained using a low level function within `copulaReg()`, which has been designed to fit GAMLSS with two or three parameter response distributions and additive predictors. An initial value for the copula parameter is obtained by using a transformation of the empirical Kendall's association between the responses.

It is worth stressing that the analytical score and Hessian of $\ell(\boldsymbol{\delta})$ required for estimation have been derived in a modular fashion. For instance, the score is defined by

$$\frac{\partial \ell(\boldsymbol{\delta})}{\partial \boldsymbol{\beta}_{\mu_1}} = \sum_{i=1}^{n} \left\{ \frac{1}{f_1(y_{1i}|\mu_{1i}, \sigma_{1i}, \nu_{1i})} \frac{\partial f_1(y_{1i}|\mu_{1i}, \sigma_{1i}, \nu_{1i})}{\partial \mu_{1i}} + \frac{1}{c\left(\mathcal{F}_1(y_{1i}|\mu_{1i}, \sigma_{1i}, \nu_{1i}), \mathcal{F}_2(y_{2i}|\mu_{2i}, \sigma_{2i}, \nu_{2i}); \theta_i\right)} \times \frac{\partial c\left(\mathcal{F}_1(y_{1i}|\mu_{1i}, \sigma_{1i}, \nu_{1i}), \mathcal{F}_2(y_{2i}|\mu_{2i}, \sigma_{2i}, \nu_{2i}); \theta_i\right)}{\partial \mathcal{F}_1(y_{1i}|\mu_{1i}, \sigma_{1i}, \nu_{1i})} \frac{\partial \mathcal{F}_1(y_{1i}|\mu_{1i}, \sigma_{1i}, \nu_{1i})}{\partial \mu_{1i}} \right\} \frac{\partial \mu_{1i}}{\partial \eta_{\mu_1 i}} \mathbf{Z}_{\mu_1 i},$$

(6)

the first derivatives of $\ell(\boldsymbol{\delta})$ with respect to $\boldsymbol{\beta}_{\mu_2}, \boldsymbol{\beta}_{\sigma_1}, \boldsymbol{\beta}_{\sigma_2}, \boldsymbol{\beta}_{\nu_1}$ and $\boldsymbol{\beta}_{\nu_2}$ (which are not reported here as their structure is very similar to (6)) and

$$\frac{\partial \ell(\boldsymbol{\delta})}{\partial \boldsymbol{\beta}_{\theta}} = \sum_{i=1}^{n} \left\{ \frac{1}{c\left(\mathcal{F}_1(y_{1i}|\mu_{1i}, \sigma_{1i}, \nu_{1i}), \mathcal{F}_2(y_{2i}|\mu_{2i}, \sigma_{2i}, \nu_{2i}); \theta_i\right)} \times \frac{\partial c\left(\mathcal{F}_1(y_{1i}|\mu_{1i}, \sigma_{1i}, \nu_{1i}), \mathcal{F}_2(y_{2i}|\mu_{2i}, \sigma_{2i}, \nu_{2i}); \theta_i\right)}{\partial \theta_i} \frac{\partial \theta_i}{\partial \eta_{\theta_i}} \right\} \mathbf{Z}_{\theta i}.$$

Looking, for instance, at equation (6), we see that there are two components which depend only on the chosen copula, three terms which are marginal distribution dependent and one derivative whose form will depend on the adopted link function between $\mu_{1i}$ and $\eta_{\mu_1 i}$. This means that it will be easy to extend our algorithm to other copulae and marginal distributions not included in Tables 1 and 2 as long as their cdfs and probability density functions are known and their derivatives with respect to their parameters exist. If a derivative is difficult and/or computationally expensive to compute then appropriate numerical approximations can be used. The score vectors and Hessian matrices for all combinations of copulae and marginal distributions considered here have been verified using the facilities available in the `numDeriv` R package (Gilbert & Varadhan, 2015).

# 4  Further details

At convergence, reliable point-wise confidence intervals for linear and non-linear functions of the model coefficients (e.g., smooth components, copula parameter, joint and conditional predicted probabilities) are obtained using $\boldsymbol{\delta} \stackrel{.}{\sim} \mathcal{N}(\hat{\boldsymbol{\delta}}, -\hat{\mathcal{H}}_p^{-1})$ where $\mathcal{H}_p$, the penalized Hessian, is defined in SM-1. The rationale for using this result is provided in Marra & Wood (2012) and references therein, whereas some examples of interval construction are given in Radice et al. (2015). To test smooth components for equality to zero, the results discussed in Wood (2013a) and Wood (2013b) are employed. Proving consistency of the proposed estimator is beyond the scope of this paper but the results in Wojtys & Marra (2015), which extend the theoretical foundation for generalized additive models established so far, could be adapted to the current context.

The proposed approach generally proved to be fast and reliable. In our experience, convergence failure typically occurs when the model is misspecified and/or the sample size is low compared to the complexity of the model. Examples of misspecification include using a Clayton copula rotated by 90 degrees when a positive association between the margins is present instead, using marginal distributions that do not fit the responses reasonably well, and employing a copula which does not accommodate the type and/or strength of dependence between the margins (e.g., using the AMH copula when the association between the margins is strong). When comparing competing models (for instance, by keeping the equations' specifications fixed and changing the copula), we observed that if the computing time for a set of alternatives is considerably higher than that of another set then this usually means that the models requiring more iterations for the algorithm to converge are not able to fit the data very well. It is also worth bearing in mind that the use of three parameter marginal distributions requires the data to be more informative as compared to a situation in which two parameter distributions are used instead. `copulaReg()` produces some warnings if there is a convergence issue, and function `conv.check()` provides some detailed diagnostics about the fitted model.

# 5  Model building

The flexibility of the proposed framework means that the researcher has to be able to choose a suitable copula function and response distributions as well as select relevant covariates in the model's additive predictors. To this end, we recommend using the Akaike information criterion (AIC) and/or Bayesian information criterion (BIC), normalized quantile residuals (Dunn & Smyth, 1996) and hypothesis testing. Since many choices need to be made, model building can become a time consuming and daunting process when working with large data sets and many candidate regressors. To facilitate the process, we suggest following roughly the guidelines of Klein et al. (2015) who argue that each of the above criteria is most useful for specific aspects of model building. In short, quantile residuals can be used to assess the goodness of fit of the marginal distributions and AIC/BIC to find a best fitting model given some pre-selected response distributions. The criteria are discussed below in more detail.

Quantile residuals for each margin are defined as $\hat{r}_{mi} = \Phi^{-1}\{\mathcal{F}_m(y_{mi}|\hat{\mu}_{mi}, \hat{\sigma}_{mi}, \hat{\nu}_{mi})\}$, for $i = 1, \ldots, n$ and $m = 1, 2$, where $\Phi^{-1}(\cdot)$ is the inverse distribution function of a standard normal distribution. If $\mathcal{F}_m(y_{mi}|\hat{\mu}_{mi}, \hat{\sigma}_{mi}, \hat{\nu}_{mi})$ is close to the true distribution then the $\hat{r}_{mi}$ follow approximately a standard normal distribution, hence a normal Q-Q plot of such residuals is a useful graphical tool for detecting lack of fit of the marginal distributions. We observed that, in practice, quantile residuals are fairly robust to the exact specification of the predictors of the distribution's parameters (this has also been found by Klein et al. (2015)). Therefore, the choice of marginal distributions can be based, for example, on more or less complex predictor specifications. Also, note that adequate marginal fits are necessary but not sufficient conditions for a satisfactory fit of the multivariate model. Functions `resp.check()` and `post.check()` in `SemiParBIVProbit` produce, respectively, histograms of the marginal response and normalized quantile residuals and normal Q-Q plots of the residuals. The former function does not account for covariates in the model and could be used prior to fitting as a rough guide to narrow down the set of plausible choices. The latter takes covariates into account and has to be used post estimation.

AIC and BIC are defined as $-2\ell(\hat{\boldsymbol{\delta}}) + 2edf$ and $-2\ell(\hat{\boldsymbol{\delta}}) + \log(n)edf$, respectively, where the log-likelihood is evaluated at the penalized parameter estimates and $edf = \text{tr}(\hat{\mathbf{A}}_{\hat{\boldsymbol{\lambda}}})$ with $\mathbf{A}$, the hat matrix, defined in SM-1. Given some marginal distributions, AIC and/or BIC can be used to select a copula function and the most relevant covariates in the model's predictors (in a stepwise backward and/or forward fashion). To favor more parsimonious models, small differences in the AIC/BIC values of competing models can be assisted by looking at the significance of the esti-mated effects; for instance, a covariate could be excluded if the respective parameter's p-value is larger than 5%. Here, the relevant R functions are `AIC()`, `BIC()`, `summary()` and `plot()`. As for the choice of copula, for empirical applications, we recommend first using Gaussian, Frank, AMH and/or FGM. Then, only the rotated versions of Clayton, Joe and/or Gumbel that are consistent with the direction (positive or negative) of the association between the margins should also be considered.

# 6 Empirical illustrations

The next sections illustrate the potential of the proposed bivariate copula additive modeling framework using two empirical case studies based on electricity and birth data, and simulated data.

## 6.1 Analysis of Spanish electricity price and demand data

The aim of this section is to build time-series models for electricity price and demand. In the engineering and econometric literature electricity demands are related with electricity prices throughout the time and one way of achieving this is via transfer function models (e.g., Nogales & Conejo, 2006). Here, we take a different approach by relating price and demand of energy using copulae. We also quantify the effect of prices of raw materials (oil, gas and coal) on electricity price and demand. We use working-daily data from January 1, 2002 to October 31, 2008 which are available from the R package `MSwM` (Sanchez-Espigares & Lopez-Moreno, 2014).

The first step is to choose the margins. Following the guidelines of Section 5, we choose the normal and Gumbel distributions for price and demand, respectively. As for the choice of copula we start off with the normal. We also allow the dependence between the margins, location and scale parameters to vary with raw material prices. In addition to these covariates, we employ a time variable as the underlying electricity prices and demands tend to vary with time, for reasons which may have little or nothing to do with material prices. When we attempt to fit a copula model in which all variables (i.e., time, oil, gas and coal prices) enter the five equations (two equations for the location parameters, two for the scale parameters and one equation for the association parameter) the algorithm fails to converge. This suggests that the sample size is perhaps low compared to the complexity of the model (see Section 4). We, therefore, try out more parsimonious specifications. Following a model building strategy along the lines of the recommendations in

Sections 4 and 5, we arrive at

```
eq.mu.1     <- Price  ~ s(t, k = 60) + s(Oil)          + s(Coal)
eq.mu.2     <- Demand ~ s(t, k = 60) + s(Oil) + s(Gas) + s(Coal)
eq.sigma2.1 <-        ~ s(t, k = 60)
eq.sigma2.2 <-        ~ s(t, k = 60) + s(Oil) + s(Gas)
eq.theta    <-        ~ s(t, k = 60)

fl <- list(eq.mu.1, eq.mu.2, eq.sigma2.1, eq.sigma2.2, eq.theta)

outN <- copulaReg(fl, margins = c("N", "GU"), data = energy, ...)
```

where the value of $60$ for the number of basis functions (`k`) for the smooths of `t` has been chosen to be a fraction (about 3.4%) of the sample size ($n = 1784$). This value implies that there are approximately 10 basis functions per year. As explained, for instance, in Peng & Dominici (2008), when `k` per year is small (say 2) only the long-term trend and seasonality will be accounted for and other sub-seasonal and shorter-term variations will remain in the data. At 10 or 12 bases per year, variation in the data longer than a timescale of about one week will be modeled; see also Wood (2006, Chapter 5). Note that the R model formula consists of a list of five equations which refer to $\eta_{\mu_1}$, $\eta_{\mu_2}$, $\eta_{\sigma_1^2}$, $\eta_{\sigma_2^2}$ and $\eta_\theta$, respectively. The first two equations always require a response whereas, to avoid redundancies, the remaining equations do not. The total number of estimated parameter is 363 and the computing time was about 12 minutes on a 2.20-GHz Intel(R) Core(TM) computer running Windows 7.

The overall Kendall's $\hat{\tau}$ and $\hat{\theta}$ are positive and significant (see `summary(outN)`), however some of the individual $\hat{\tau}$ and $\hat{\theta}$ assume negative values. This suggests that copulae which cannot account for positive and negative dependencies at the same time should not be used. Therefore, the remaining choices are `F`, `AMH` and `FGM` where the last two can only account for weak dependencies ($-0.18 \leq \tau \leq 0.33$ and $-0.22 \leq \tau \leq 0.22$, respectively). When trying out these alternatives, the computing time increased considerably as the algorithm required a higher number of iterations to reach convergence. As explained in Section 4, this may occur when a model is not able to fit the data very well; this is supported by the AIC and BIC values for the four models which indicate that the normal copula provides the best fit to the data. Marginal residual plots for the final model are shown in Figure 1 and suggest that the choice of marginal distributions is reasonable.

Using the fitted model, we build the plot in Figure 2 which shows that the correlation between `Price` and `Demand` fluctuates around $0.5$ (a similar plot could be produced for Kendall's $\tau$).

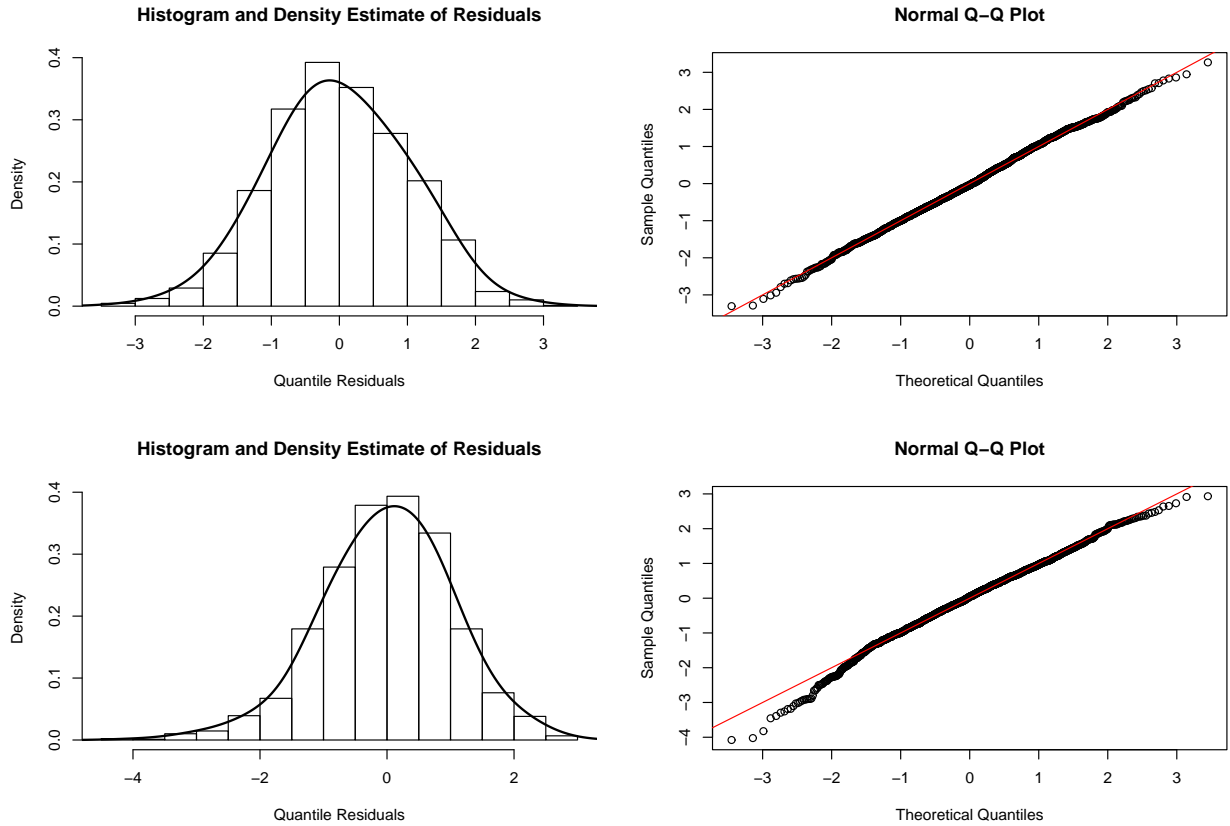

Figure 1: Histograms of normalized quantile residuals for electricity price (top) and demand (bottom) and normal Q-Q plots of residuals. These have been produced after fitting a Gaussian copula model with normal and Gumbel margins to electricity price and demand data.

Many of the intervals do not contain zero: after accounting for raw material prices, a significant association between the two responses which varies over time still persists. Moving on to the covariate effects and focusing, for instance, on the first equation, Figure 3 displays the impacts of t, Oil and Coal on Price. The plots show a cyclic trend with maximum and minimum peaks and suggest that on average electricity price tends to increase with Oil, and decrease and then stabilize with Coal. Figure 4 reports the estimates and intervals for $\sigma_1^2$ suggesting that the variability of Price is not constant over time. We could also predict joint and conditional probabilities of interest from the model. This point is illustrated in the next section.

The R code used for the above analysis is given in SM-2.

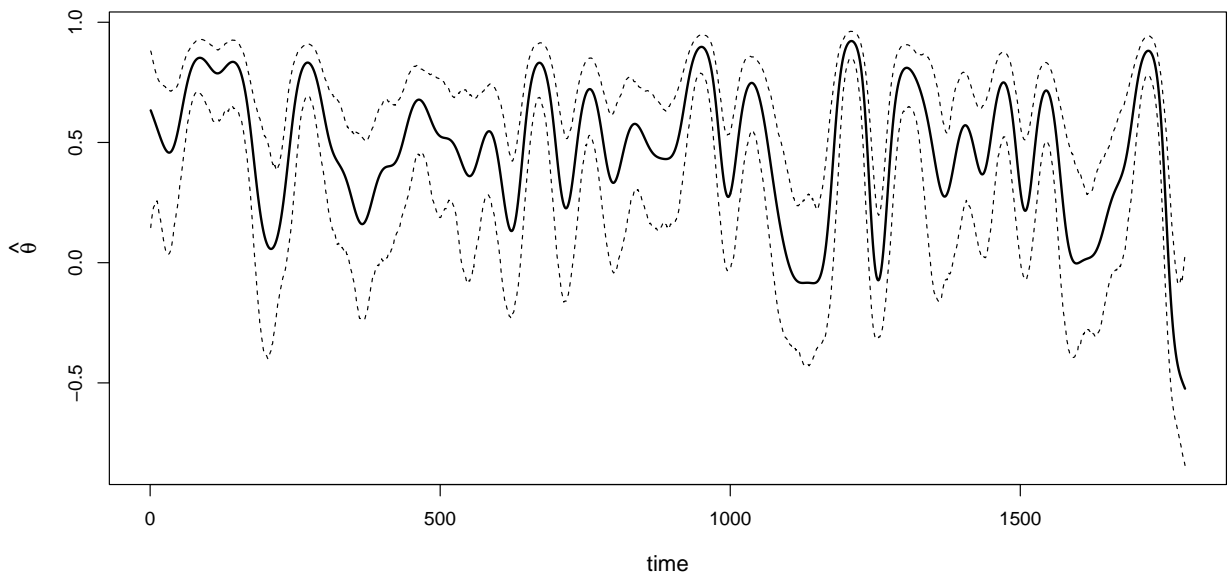

Figure 2: Estimates and 95% intervals for $\theta$ over time from a Gaussian copula model with normal and Gumbel margins fitted to electricity price and demand data.

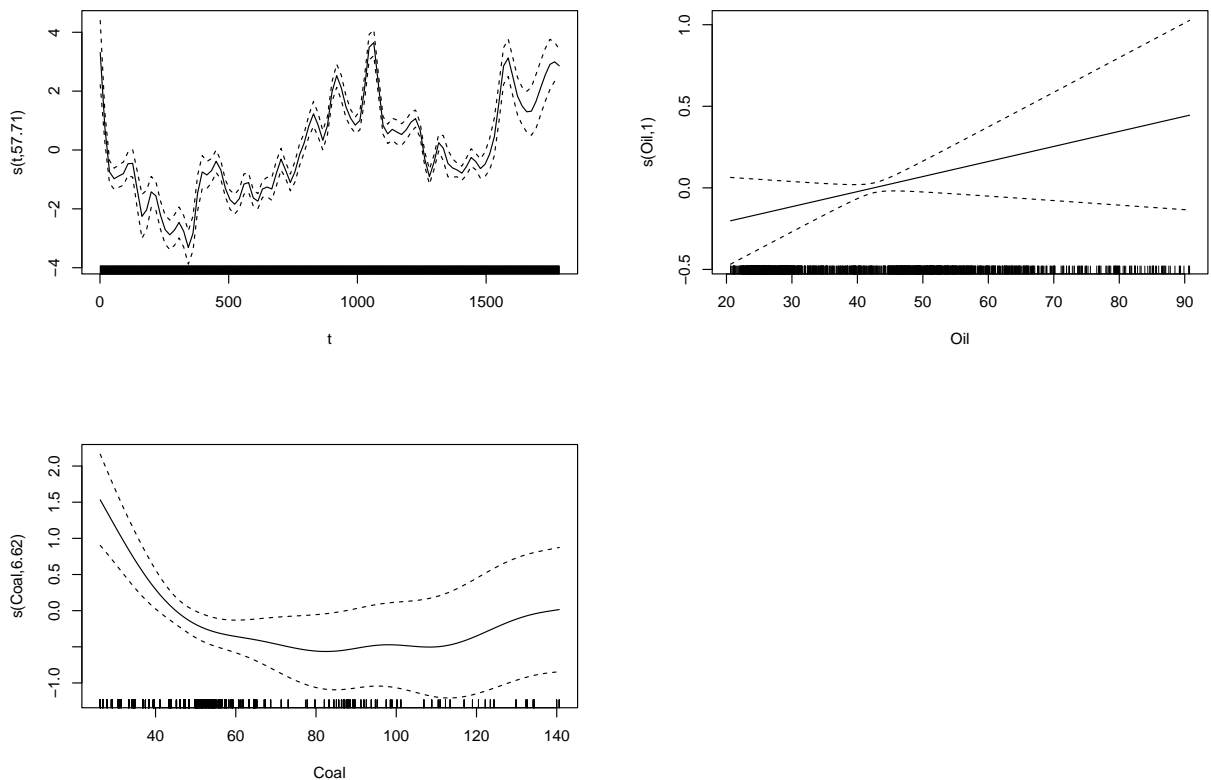

Figure 3: Estimated smooth effects of time, oil and coal prices on electricity price and associated 95% point-wise intervals obtained when fitting a Gaussian copula model with normal and Gumbel margins to electricity price and demand data. The jittered rug plot, at the bottom of each graph, shows the covariate values. The number in brackets in the y-axis caption represents the effective degrees of freedom (edf) of the smooth curve (see SM-1 for the definition of edf). Note that the estimated smooth functions are centered around zero because of centering identifiability constraints (see Section 2.1).

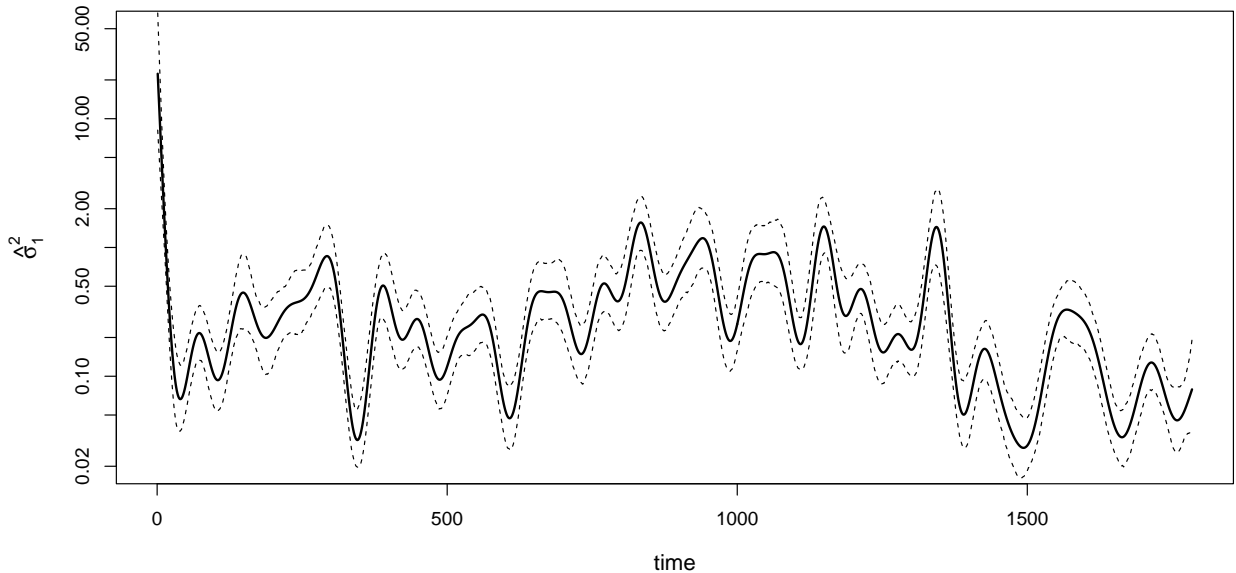

Figure 4: Estimates and 95% intervals for $\sigma_1^2$, the variance for electricity price, over time from a Gaussian copula model with normal and Gumbel margins fitted to electricity price and demand data. Results are reported on a log-scale for the y-axis.

## 6.2 North Carolina birth data analysis

The analysis in this section uses 2010 birth data from the North Carolina Center for Health Statistics (http://www.schs.state.nc.us/) which provides details on all live births occurred within the State of North Carolina, including information on infant and maternal health and parental characteristics. The data cover maternal demographic information, pregnancy related events and outcomes, maternal medical complications, newborn conditions and maternal health behaviors. The choice of variables largely follows the work by Neelon et al. (2012) and the analysis reported below is for female infants (similar results were obtained for male infants). The responses are birth weight in grams (bwgram) and gestational age in weeks (wksgest). The covariates are maternal ethnicity (nonhisp, categorized as non-Hispanic and Hispanic), singleton birth (multbirth, born as a multiple or single birth), maternal age (mage in years), mother's marital status (married) and county (county, indicating the North Carolina county of residence of the mother).

The goal is to build a spatial bivariate copula regression model for the joint analysis of bwgram and wksgest. As highlighted in the previous section, the proposed modeling framework is appealing as it allows for flexible joint and marginal inferences. In this case, the bivariate copula model can be used to estimate the association (adjusted for covariates) between bwgram and wksgest by county, to quantify the effects of covariates on bwgram and wksgest, and to calculate joint and conditional probabilities of interest.

We first analyze the marginal distributions of bwgram and wksgest. We find that the choice of the marginal distribution is fairly non-sensitive to the exact specification of additive predictor. The normal Q-Q plots of the normalized quantile residuals and AIC consistently suggest that the best fits for bwgram and wksgest are achieved using the logistic and Gumbel distributions, respectively. We then proceed by fitting bivariate models for bwgram and wksgest following the approach adopted in the previous section which is based on the guidelines outlined in Section 5. The final model is

```
eq.mu.1     <- bwgram  ~ nonhisp + multbirth + married + s(mage) +
                         s(county, bs = "mrf", xt = xt)
eq.mu.2     <- wksgest ~ nonhisp + multbirth + married + s(mage) +
                         s(county, bs = "mrf", xt = xt)
eq.sigma2.1 <-         ~ nonhisp + multbirth + married + s(mage) +
                         s(county, bs = "mrf", xt = xt)
eq.sigma2.2 <-         ~           multbirth + married + s(mage) +
                         s(county, bs = "mrf", xt = xt)
eq.theta    <-         ~ nonhisp + multbirth           + s(mage) +
                         s(county, bs = "mrf", xt = xt)

fl <- list(eq.mu.1, eq.mu.2, eq.sigma2.1, eq.sigma2.2, eq.theta)

outC0 <- copulaReg(fl, margins = c("LO", "GU"), BivD = "C0",
                   data = datNC, ...)
```

where a Clayton copula is used to join the logistic and Gumbel distributions for the two responses. The first two equations refer to the $\mu$ parameters of bwgram and wksgest, the third and fourth to the $\sigma^2$ parameters and the last to $\theta$. All parameters are modeled using predictors involving factor, continuous and regional variables. The use of mrf smoothers in all equations ensures that the distribution parameters vary smoothly across counties. The total number of observations and estimated parameters are 56940 and 558, respectively, and the computing time was about 25 minutes.

An analysis similar to that produced in Section 6.1, showing for instance some estimated smooth function and Kendall's $\hat{\tau}$ by county is given in SM-4 to save space. Figure 5 shows the joint probabilities of low weight birth babies and premature deliveries in North Carolina when using a copula model and an independence model (which assumes that bwgram an wksgest are not associated after accounting for covariates). This joint probability was calculated for all the observations in the dataset and then averaged by county. The AIC and BIC values for the copula and independence models suggest that the former provides a better fit to the data. As it can be seen from Figure 5, in this case, assuming independence leads to underestimated probabilities.

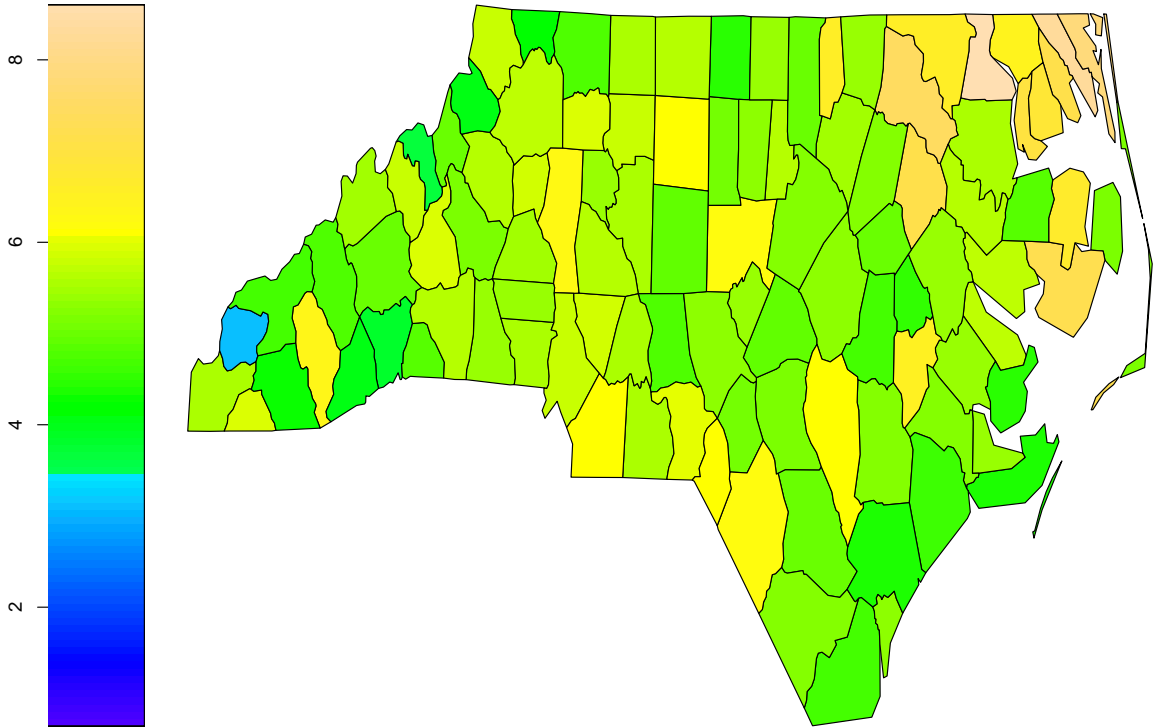

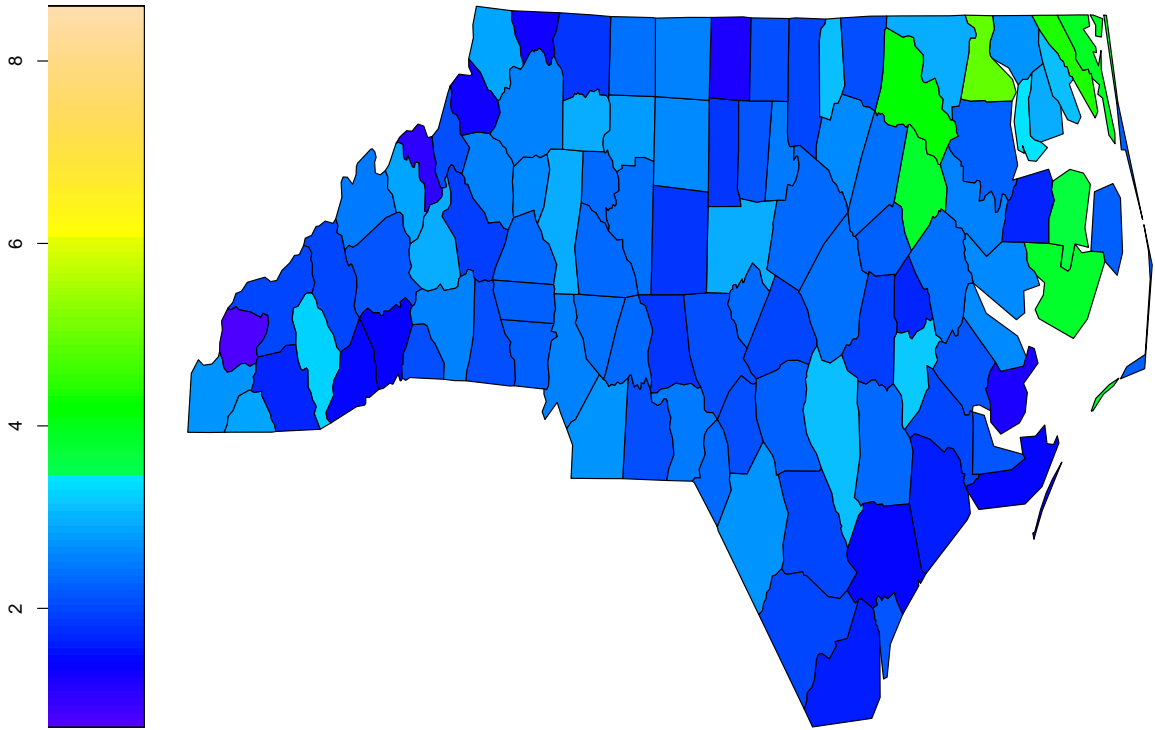

Figure 5: Joint probabilities that `bwgram` is less than or equal to 2500 grams and that `wksgest` is less than or equal to 37 weeks by county in North Carolina. These have been calculated using a Clayton copula model and an independence model (assuming that `bwgram` and `wksgest` are not associated after accounting for covariates).

The R code used for the above analysis is given in SM-3.

## 6.3 Simulated data example

To assess the empirical effectiveness of the proposed methodology in a controlled setting, we conducted a simulation study.

We created two continuous outcomes, two continuous regressors and a binary covariate (denoted as $y_1$, $y_2$, $x_1$, $x_2$, and $x_3$, respectively). The two responses were assumed to follow inverse Gaussian and Singh-Maddala distributions, respectively. Variables $x_1$, $x_2$ and $x_3$ were generated from correlated uniform distributions over $[0, 1]$ (Gentle, 2003). Variable $x_3$ was then dichotomized so that each value had a $50\%$ chance of appearing. The two responses, $y_1$ and $y_2$, were joined using a Joe copula. Linear and non-linear effects of the regressors on the parameters of the inverse Gaussian and Singh-Maddala distributions as well as copula parameter were introduced. The predictor specification used for $\theta$ yielded an average copula parameter value of $7.9$, which corresponds to a Kendall's $\tau$ of $0.76$. Sample sizes were set to 1000 and 2000, and the number of replicates to 250. The copulae employed were J0 (the correct model), J180, C0, C180, G0, G180, F and N. We did not consider AMH and FGM as their dependence coverages do not include the above Kendall's $\tau$ value. The other rotated versions of Joe, Clayton and Gumbel were also not considered as they are not consistent with the simulated positive associations between the margins. The R code used to simulate the data is given in SM-5.

For each replicate and fitted model, we stored the estimated linear effects, AIC, BIC and estimated smooth functions evaluated at 200 fixed values in the ranges of the respective covariates. Figure 6 shows that the estimated curves recover the underlying functions fairly well. There are some exceptions, especially for $n = 1000$, where the estimated functions are either wigglier or smoother than they should be. This does not come as a surprise and has vanishing probability for increasing sample size (e.g., Radice et al., 2015; Yoshida & Naito, 2014). Figure 7 shows the results for the parametric effects; the performance of the estimator is satisfactory and improves as the sample size grows large.

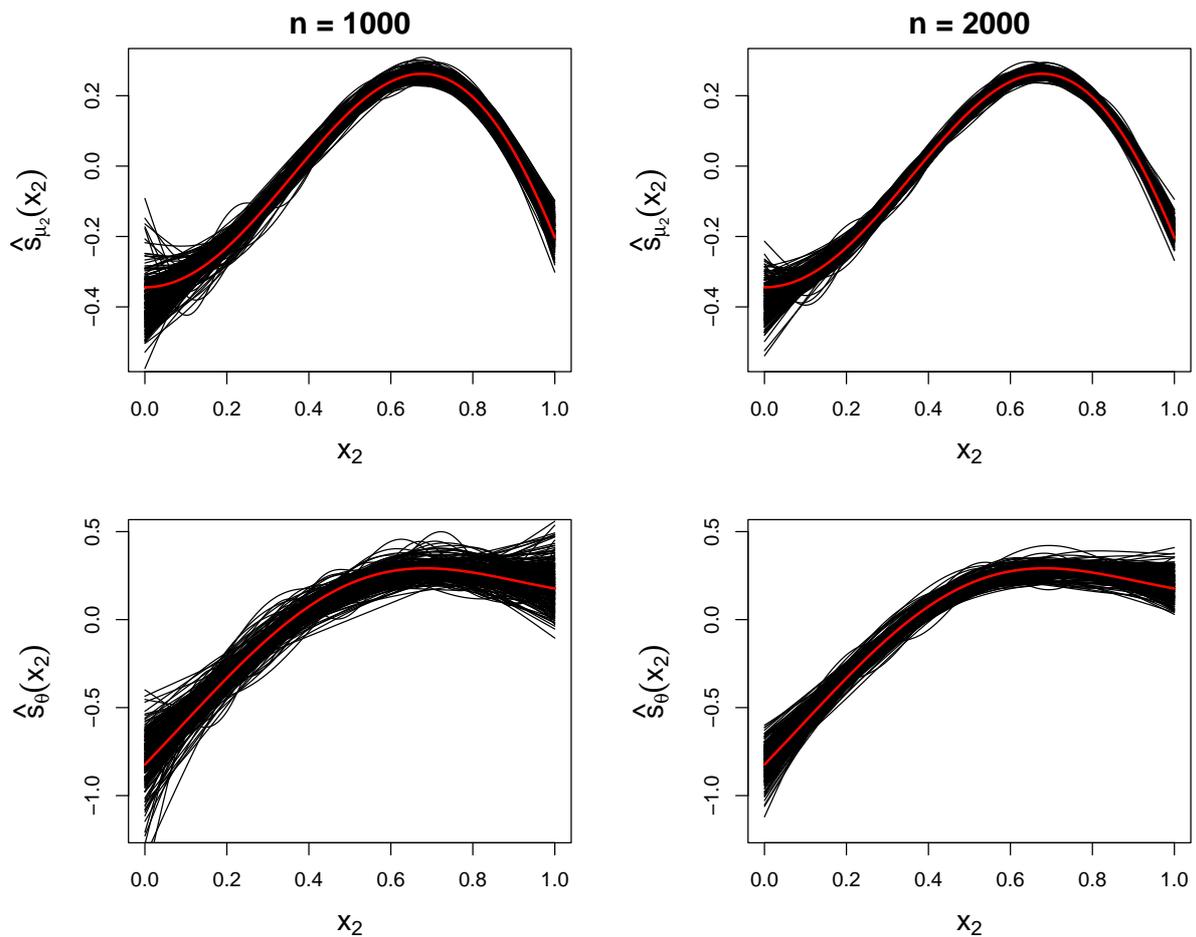

Figure 6: Estimated smooth functions for $s_{\mu_2}(x_2)$ and $s_\theta(x_2)$ obtained by fitting Joe copula models to simulated data. The plots in the first row are the effects of $x_2$ on the $\mu$ parameter of the Singh-Maddala distribution on the scale of the additive predictor, whereas those in the second row are the effects of $x_2$ on the $\theta$ parameter of the Joe copula on the scale of the predictor. The black lines in each plot represent the estimated smooth functions from all 250 replicates, evaluated at 200 fixed values in $[0, 1]$. The true functions are represented by the red solid lines. The notation used in the y-axis labels is consistent with that used in SM-5. More details are given in Section 6.3.

The proportions of times that the models were selected by AIC and BIC over the replicates were also calculated. For $n = 1000$, the only selected models were J0 (the correct model) and C180 with proportions of $0.61$ and $0.39$ when using AIC, and of $0.59$ and $0.41$ when using BIC. For $n = 2000$, the proportions were $0.65$ and $0.35$ when using AIC, and $0.63$ and $0.37$ when using BIC. It is not surprising that C180 competed with J0 as these are the two most similar copulae considered in the simulations.

# 7  Discussion

We have introduced a modeling framework for bivariate copula additive models for location, scale and shape. The modularity of the estimation approach allows for easy inclusion of potentially any parametric continuous marginal distribution and one-parameter copula function as long as the cdfs and probability density functions are known and their derivatives with respect to their parameters exist. Parameter estimation is carried out within a penalized maximum likelihood estimation framework with integrated automatic multiple smoothing parameter selection, and known and reliable inferential results from the smoothing literature are employed for interval construction and hypothesis testing. The proposed models can be easily used via `copulaReg()` in `SemiParBIVProbit` and the potential of the approach has been demonstrated using real data analyzes and simulated data.

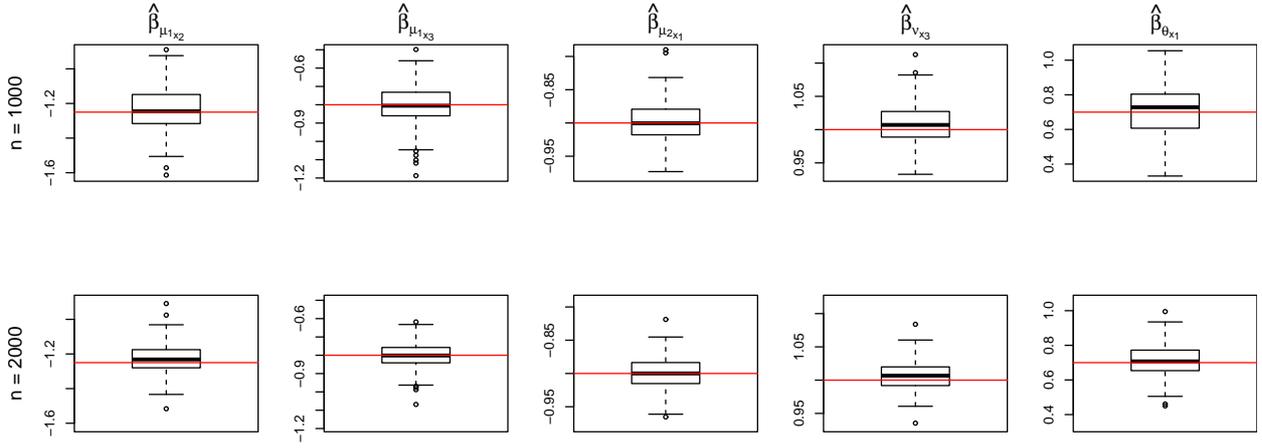

Figure 7: Boxplots of estimated parametric effects for $\beta_{\mu_1,x_2}$, $\beta_{\mu_1,x_3}$, $\beta_{\mu_2,x_1}$, $\beta_{\nu,x_3}$ and $\beta_{\theta,x_1}$ obtained by fitting Joe copula models to data simulated as described in Section 6.3. The true values are represented by the red horizontal lines. The notation used in the main labels is consistent with that used in SM-5.

Future releases of `SemiParBIVProbit` will incorporate more copulae and marginal distributions as well as facilities for comparing the predictive ability of competing models based, for instance, on proper scoring rules (Gneiting & Raftery, 2007).

Future research will look into the feasibility of strengthening the framework described in this article by incorporating two-parameter and non-exchangeable copulae (e.g., Durante, 2009; Frees & Valdez, 1998; Brechmann & Schepsmeier, 2013). In both cases, extra distribution parameters will have to be estimated and it is not yet clear whether the resulting likelihood function will be informative "enough" to allow for reliable parameter estimation in the presence of flexible covariate effects. Another interesting extension would be to consider systems involving more than two responses using C- and D-Vine copulae (e.g., Brechmann & Schepsmeier, 2013).

# References

Acar, E. F., Craiu, V. R., & Yao, F. (2013). Statistical testing of covariate effects in conditional copula models. *Electronic Journal of Statistics*, (pp. 2822–2850).

Brechmann, E. C. & Schepsmeier, U. (2013). Modeling dependence with c- and d-vine copulas: The R package CDVine. *Journal of Statistical Software*, 52(3), 1–27.

Cherubini, U., Luciano, E., & Vecchiato, W. (2004). *Copula Methods in Finance*. Wiley, London.

Craiu, V. R. & Sabeti, A. (2012). In mixed company: Bayesian inference for bivariate conditional copula models with discrete and continuous outcomes. *Journal of Multivariate Analysis*, 110, 106–120.

Dunn, P. K. & Smyth, G. K. (1996). Randomized quantile residuals. *Journal of Computational and Graphical Statistics*, 5, 236–245.

Durante, F. (2009). Construction of non-exchangeable bivariate distribution functions. *Statistical Papers*, 50, 383–391.

Eilers, P. H. C. & Marx, B. D. (1996). Flexible smoothing with B-splines and penalties. *Statistical Science*, 11(2), 89–121.

Frees, E. W. & Valdez, E. A. (1998). Understanding relationships using copulas. *North American Actuarial Journal*, 2, 1–25.

Gentle, J. E. (2003). *Random number generation and Monte Carlo methods*. Springer-Verlag, London.

Gijbels, I., Veraverbeke, N., & Omelka, M. (2011). Conditional copulas, association measures and their applications. *Computational Statistics and Data Analysis*, 55, 1919–1932.

Gilbert, P. & Varadhan, R. (2015). *numDeriv: Accurate Numerical Derivatives*. R package version 2014.2-1.

Gneiting, T. & Raftery, A. E. (2007). Strictly proper scoring rules, prediction, and estimation. *Journal of the American Statistical Association*, 102, 359–378.

Hastie, T. J. & Tibshirani, R. J. (1990). *Generalized Additive Models*. Chapman & Hall/CRC, London.

Kauermann, G., Krivobokova, T., & Fahrmeir, L. (2009). Some asymptotics results on generalized penalized spline smoothing. *Journal of Royal Statistical Society Series B*, 71, 487–503.

Kauermann, G., Schellhase, C., & Ruppert, D. (2013). Flexible copula density estimation with penalized hierarchical b-splines. *Scandinavian Journal of Statistics*, 40, 685–705.

Klein, N. & Kneib, T. (2015). Simultaneous inference in structured additive conditional copula regression models: a unifying bayesian approach. *Statistics and Computing*.

Klein, N., Kneib, T., Klasen, S., & Lang, S. (2015). Bayesian structured additive distributional regression for multivariate responses. *Journal of the Royal Statistical Society, Series C*, 64, 569–591.

Kolev, N. & Paiva, D. (2009). Copula-based regression models: A survey. *Journal of Statistical Planning and Inference*, 139, 3847–3856.

Kraemer, N. & Silvestrini, D. (2015). *CopulaRegression: Bivariate Copula Based Regression Models*. R package version 0.1-5.

Kramer, N., Brechmann, E. C., Silvestrini, D., & Czado, C. (2012). Total loss estimation using copula-based regression models. *Insurance: Mathematics and Economics*, 53, 829–839.

Lambert, P. (2007). Archimedean copula estimation using bayesian splines smoothing techniques. *Computational Statistics and Data Analysis*, 51, 6307–6320.

Marra, G. & Radice, R. (2016). *SemiParBIVProbit: Semiparametric Copula Bivariate Probit Modelling*. R package version 3.7.

Marra, G. & Wood, S. (2012). Coverage properties of confidence intervals for generalized additive model components. *Scandinavian Journal of Statistics*, 39, 53–74.

Neelon, B., Anthopolos, R., , & Miranda, M. L. (2012). A spatial bivariate probit model for correlated binary data with application to adverse birth outcomes. *Statistical Methods in Medical Research*, 23(2), 119–133.

Nelsen, R. (2006). *An Introduction to Copulas*. New York: Springer.

Nogales, F. J. & Conejo, A. J. (2006). Electricity price forecasting through transfer function models. *Journal of the Operational Research Society*, 57, 350–356.

Peng, R. D. & Dominici, F. (2008). *Statistical Methods for Environmental Epidemiology with R: A Case Study in Air Pollution and Health*. Springer.

Radice, R., Marra, G., & Wojtys, M. (2015). Copula regression spline models for binary outcomes. *Statistics and Computing*, (pp. 1–15).

Rigby, R. A. & Stasinopoulos, D. M. (2005). Generalized additive models for location, scale and shape (with discussion). *Journal of the Royal Statistical Society, Series C*, 54, 507–554.

Rue, H. & Held, L. (2005). *Gaussian Markov Random Fields*. New Haven: Chapman & Hall/CRC, Boca Raton, FL.

Ruppert, D., Wand, M. P., & Carroll, R. J. (2003). *Semiparametric Regression*. Cambridge University Press, New York.

Sanchez-Espigares, J. A. & Lopez-Moreno, A. (2014). *MSwM: Fitting Markov Switching Models*. R package version 1.2.

Segers, J., van den Akker, R., & Werker, B. J. M. (2014). Linear b-spline copulas with applications to nonparametric estimation of copulas. *Annals of Statistics*, 42, 1911–1940.

Shen, X., Zhu, Y., & Song, L. (2008). Linear b-spline copulas with applications to nonparametric estimation of copulas. *Computational Statistics and Data Analysis*, 52, 3806–3819.

Sklar, A. (1959). Fonctions de répartition é n dimensions et leurs marges. *Publications de l'Institut de Statistique de l'Université de Paris*, 8, 229–231.

Sklar, A. (1973). Random variables, joint distributions, and copulas. *Kybernetica*, 9, 449–460.

Trivedi, P. K. & Zimmer, D. M. (2006). *Copula Modelling in Econometrics: Introduction to Practitioners*.

Wojtys, M. & Marra, G. (2015). Copula based generalized additive models with non-random sample selection. *arXiv:1508.04070*.

Wood, S. N. (2003). Thin plate regression splines. *Journal of the Royal Statistical Society Series B*, 65, 95–114.

Wood, S. N. (2006). *Generalized Additive Models: An Introduction With R*. Chapman & Hall/CRC, London.

Wood, S. N. (2013a). On p-values for smooth components of an extended generalized additive model. *Biometrika*, 100, 221–228.

Wood, S. N. (2013b). A simple test for random effects in regression models. *Biometrika*, 100, 1005–1010.

Wood, S. N. (2016). *mgcv: Mixed GAM Computation Vehicle with GCV/AIC/REML Smoothness Estimation*. R package version 1.8-11.

Yan, J. (2007). Enjoy the joy of copulas: With a package copula. *Journal of Statistical Software*, 21, 1–21.

Yee, T. W. (2016). *VGAM: Vector Generalized Linear and Additive Models*. R package version 1.0-0.

Yoshida, T. & Naito, K. (2014). Asymptotics for penalized splines in generalized additive models. *Journal of Nonparametric Statistics*, 26, 269–289.

Zimmer, D. M. & Trivedi, P. K. (2006). Using trivariate copulas to model sample selection and treatment effects: application to family health care demand. *Journal of Business & Economic Statistics*, 24, 63–76.

# Supplementary material for "A Bivariate Copula Additive Model for Location, Scale and Shape"

Tuesday 5th April, 2016

## 1 Sketch of algorithm

Estimation of $\boldsymbol{\delta}$ and $\boldsymbol{\lambda}$ is carried out in a two-step fashion:

**step 1** Holding the smoothing parameter vector fixed at $\boldsymbol{\lambda}^{[a]}$ and for a given parameter vector value $\boldsymbol{\delta}^{[a]}$, we seek to maximize equation (5) in the paper using a trust region algorithm. That is,

$$\min_{\mathbf{p}} \; \breve{\ell}_p(\boldsymbol{\delta}^{[a]}) \stackrel{\text{def}}{=} -\left\{\ell_p(\boldsymbol{\delta}^{[a]}) + \mathbf{p}^\mathsf{T}\mathbf{g}_p^{[a]} + \frac{1}{2}\mathbf{p}^\mathsf{T}\boldsymbol{\mathcal{H}}_p^{[a]}\mathbf{p}\right\} \; \text{ so that } \; \|\mathbf{p}\| \leq r^{[a]},$$

$$\boldsymbol{\delta}^{[a+1]} = \arg\min_{\mathbf{p}} \; \breve{\ell}_p(\boldsymbol{\delta}^{[a]}) + \boldsymbol{\delta}^{[a]},$$

where $a$ is an iteration index, $\mathbf{g}_p^{[a]} = \mathbf{g}^{[a]} - \mathbf{S}_{\boldsymbol{\lambda}^{[a]}}\boldsymbol{\delta}^{[a]}$ and $\boldsymbol{\mathcal{H}}_p^{[a]} = \boldsymbol{\mathcal{H}}^{[a]} - \mathbf{S}_{\boldsymbol{\lambda}^{[a]}}$. Vector $\mathbf{g}^{[a]}$ consists of $\mathbf{g}_{\mu_1}^{[a]} = \partial\ell(\boldsymbol{\delta})/\partial\boldsymbol{\beta}_{\mu_1}|_{\boldsymbol{\beta}_{\mu_1}=\boldsymbol{\beta}_{\mu_1}^{[a]}}, \ldots, \mathbf{g}_{\theta}^{[a]} = \partial\ell(\boldsymbol{\delta})/\partial\boldsymbol{\beta}_{\theta}|_{\boldsymbol{\beta}_{\theta}=\boldsymbol{\beta}_{\theta}^{[a]}}$, the Hessian matrix has elements $\boldsymbol{\mathcal{H}}_{r,h}^{[a]} = \partial^2\ell(\boldsymbol{\delta})/\partial\boldsymbol{\beta}_r\partial\boldsymbol{\beta}_h^\mathsf{T}|_{\boldsymbol{\beta}_r=\boldsymbol{\beta}_r^{[a]},\boldsymbol{\beta}_h=\boldsymbol{\beta}_h^{[a]}}$, where $r, h = \mu_1, \mu_2, \sigma_1, \sigma_2, \nu_1, \nu_2, \theta$, $\|\cdot\|$ denotes the Euclidean norm and $r^{[a]}$ is the radius of the trust region which is adjusted through the iterations. See, e.g., Geyer (2015) for full details.

**step 2** For a given smoothing parameter vector value $\boldsymbol{\lambda}^{[a]}$ and holding the main parameter vector value fixed at $\boldsymbol{\delta}^{[a+1]}$, solve the problem

$$\boldsymbol{\lambda}^{[a+1]} = \arg\min_{\boldsymbol{\lambda}} \; \mathcal{V}(\boldsymbol{\lambda}) \stackrel{\text{def}}{=} \|\mathbf{z}^{[a+1]} - \mathbf{A}_{\boldsymbol{\lambda}^{[a]}}^{[a+1]}\mathbf{z}^{[a+1]}\|^2 - \breve{n} + 2\mathrm{tr}(\mathbf{A}_{\boldsymbol{\lambda}^{[a]}}^{[a+1]}), \qquad (1)$$

where, after defining $\boldsymbol{\mathcal{I}}^{[a+1]}$ as $-\boldsymbol{\mathcal{H}}^{[a+1]}$, $\mathbf{z}^{[a+1]} = \sqrt{\boldsymbol{\mathcal{I}}^{[a+1]}}\boldsymbol{\delta}^{[a+1]} + \sqrt{\boldsymbol{\mathcal{I}}^{[a+1]}}^{-1}\mathbf{g}^{[a+1]}$, $\mathbf{A}_{\boldsymbol{\lambda}^{[a]}}^{[a+1]} =$ $\sqrt{\mathcal{I}^{[a+1]}} \left( \mathcal{I}^{[a+1]} + \mathbf{S}_{\boldsymbol{\lambda}^{[a]}} \right)^{-1} \sqrt{\mathcal{I}^{[a+1]}}$, tr($\mathbf{A}^{[a+1]}_{\boldsymbol{\lambda}^{[a]}}$) represents the number of effective degrees of freedom (edf) of the penalized model and $\check{n} = 7n$ (if a three parameter distribution is employed for both margins). Problem (1) is solved using the automatic efficient and stable approach by Wood (2004).

The two steps are iterated until the algorithm satisfies the criterion $\frac{|\ell(\boldsymbol{\delta}^{[a+1]}) - \ell(\boldsymbol{\delta}^{[a]})|}{0.1 + |\ell(\boldsymbol{\delta}^{[a+1]})|} < 1e - 07$. The use of a trust region algorithm in step 1 and of (1) in step 2 are justified in Marra et al. (2016) and Radice et al. (2015). It is worth remarking that a trust region approach is generally more stable and faster than its line-search counterparts (such as Newton-Raphson), particularly for functions that are, for example, non-concave and/or exhibit regions that are close to flat (Nocedal & Wright, 2006, Chapter 4). Note that since $\mathcal{H}$ and $\mathbf{g}$ are obtained as a byproduct of the estimation step for $\boldsymbol{\delta}$, little computational effort is required to set up the quantities required for the smoothing step.

## 2   R code for electricity price and demand data analysis

```
library(SemiParBIVProbit)
data("energy", package = "MSwM") # where energy data are from
energy$t <- 1:dim(energy)[1] # create time variable

eq.mu.1     <- Price  ~ s(t, k = 60) + s(Oil)          + s(Coal)
eq.mu.2     <- Demand ~ s(t, k = 60) + s(Oil) + s(Gas) + s(Coal)
eq.sigma2.1 <-        ~ s(t, k = 60)
eq.sigma2.2 <-        ~ s(t, k = 60) + s(Oil) + s(Gas)
eq.theta    <-        ~ s(t, k = 60)

fl <- list(eq.mu.1, eq.mu.2, eq.sigma2.1, eq.sigma2.2, eq.theta)

# Note: Gaussian copula model is the quickest (around 12 minutes)
# the other models are considerably slower as
# explained in Section 6.1 of the main paper

outN   <- copulaReg(fl, margins = c("N", "GU"),
                    data = energy, gamlssfit = TRUE)
outF   <- copulaReg(fl, margins = c("N", "GU"), BivD = "F",
                    data = energy, gamlssfit = TRUE)
outAMH <- copulaReg(fl, margins = c("N", "GU"), BivD = "AMH",
                    data = energy, gamlssfit = TRUE)
outFGM <- copulaReg(fl, margins = c("N", "GU"), BivD = "FGM",
                    data = energy, gamlssfit = TRUE)
```

```
conv.check(outN)
conv.check(outF)
conv.check(outAMH)
conv.check(outFGM)

post.check(outN)

AIC(outN, outF, outAMH, outFGM)
BIC(outN, outF, outAMH, outFGM)

ss <- summary(outN, n.sim = 1000)
# n.sim is the no. of simulated coefficient vectors from the posterior
# distribution of the estimated model parameters, which are used to
# calculate intervals for non-linear functions of the model parameters

ss

# estimated tau and theta for each observation
outN$tau
outN$theta

CItheta <- ss$CItheta
par(mar = c(5, 5, 4, 2) + 0.1)
plot(energy$t, outN$theta, type = "l", ylim = c(-0.85, 0.97),
     ylab = expression(hat(theta)),
     xlab = "time", lwd = 2, cex.lab = 1.2)
lines(energy$t, CItheta[,1], lty = 2)
lines(energy$t, CItheta[,2], lty = 2)

plot(outN, eq = 1, pages = 1, scale = 0, seWithMean = TRUE)

CIsigma21 <- ss$CIsig21
par(mar = c(5, 5, 4, 2) + 0.1)
plot(energy$t, outN$sigma21, type = "l",
     ylim = c(0.02, 50), log = "y",
     ylab = expression(hat(sigma)[1]^2),
     xlab = "time", lwd = 2, cex.lab = 1.2)
lines(energy$t, CIsigma21[, 1], lty = 2)
lines(energy$t, CIsigma21[, 2], lty = 2)
```

# 3 R code for North Carolina birth data analysis

```
library(SemiParBIVProbit)

load("datNC.RData")
# datNC.RData is available through the Journal website
# It contains two files:
# 1. the data set: datNC
# 2. Polygon shape file to build maps and mrf smoother: NC.polys

# Create list of polygons suitable for mrf smoother
xt <- list(polys = NC.polys)

eq.mu.1     <- bwgram  ~ nonhisp + multbirth + married + s(mage) +
                         s(county, bs = "mrf", xt = xt)
eq.mu.2     <- wksgest ~ nonhisp + multbirth + married + s(mage) +
                         s(county, bs = "mrf", xt = xt)
eq.sigma2.1 <-         ~ nonhisp + multbirth + married + s(mage) +
                         s(county, bs = "mrf", xt = xt)
eq.sigma2.2 <-         ~           multbirth + married + s(mage) +
                         s(county, bs = "mrf", xt = xt)
eq.theta    <-         ~ nonhisp + multbirth           + s(mage) +
                         s(county, bs = "mrf", xt = xt)

fl <- list(eq.mu.1, eq.mu.2, eq.sigma2.1, eq.sigma2.2, eq.theta)

outC0 <- copulaReg(fl, margins = c("LO", "GU"), BivD = "C0",
                   data = datNC, gc.l = TRUE, gamlssfit = TRUE)
conv.check(outC0)

# Comparison of AICs/BICs for copula and independence models
AIC(outC0)
AIC(outC0$gamlss1$lk) + AIC(outC0$gamlss2$lk)
BIC(outC0)
BIC(outC0$gamlss1$lk) + BIC(outC0$gamlss2$lk)

lr <- length(NC.polys)
pr.jointC <- pr.indepC <- NA
sigma21 <- tau <- NA

# jc.probs calculates joint or conditional probabilities of interest
# it also delivers intervals if intervals = TRUE
pr.joint <- jc.probs(outC0, 2500, 37)$p12*100
pr.indep <- jc.probs(outC0, 2500, 37, type = "independence")$p12*100
```

```
for(i in 1:lr){
   pr.jointC[i] <- mean(pr.joint[datNC$county==i])
   pr.indepC[i] <- mean(pr.indep[datNC$county==i])
   tau[i]       <- mean(outC0$tau[datNC$county==i])
   sigma21[i]   <- mean(outC0$sigma21[datNC$county==i])
             }

polys.map(NC.polys, pr.jointC, rev.col = FALSE,
          main = "Joint probabilities (in %) from copula model",
          scheme = "topo", cex.main = 1.7, zlim = c(0.7, 8.6))

polys.map(NC.polys, pr.indepC, rev.col = FALSE, cex.main = 1.7,
          main = "Joint probabilities (in %) from independence model",
          scheme = "topo", zlim = c(0.7, 8.6))

ss <- summary(outC0)
ss

plot(outC0, eq = 1, select = 1, scale = 0, jit = TRUE, cex.lab = 1.3)

polys.map(NC.polys, sqrt(pi^2*sigma21/3), rev.col = FALSE,
          main = expression(paste(hat(sigma)," of bwgram")),
          scheme = "topo", cex.main = 1.7)

polys.map(NC.polys, tau, rev.col = FALSE, cex.main = 1.7,
          main = expression(paste("Kendall's ",hat(tau))),
          scheme = "topo")

# 95% intervals for Kendall's tau for the first 10 observations
ss$CIkt[1:10,]
```

# 4 Further results from North Carolina birth data analysis

Focusing, for example, on the first equation, the estimated smooth effect of mage on bwgram, reported in Figure 1, shows that on average birth weight tends to increase with mother's age (with a pick at 36-37 years of age) and then to decrease. From an inferential point of view there is bigger uncertainty associated with the curve estimate for women older than 40 years old; this is due to the sparsity of the regressor. Note that the estimated smooth function is centered around zero because of centering (identifiability) constraints (see Section 2.1 in the main paper), however this does not affect interpretation. There is also presence of heteroscedasticity in birth weight; Figure 2 (top plot) shows the estimated standard deviation of `bwgram` averaged by `nonhisp`, `multbirth`, `married` and `mage` within each county.

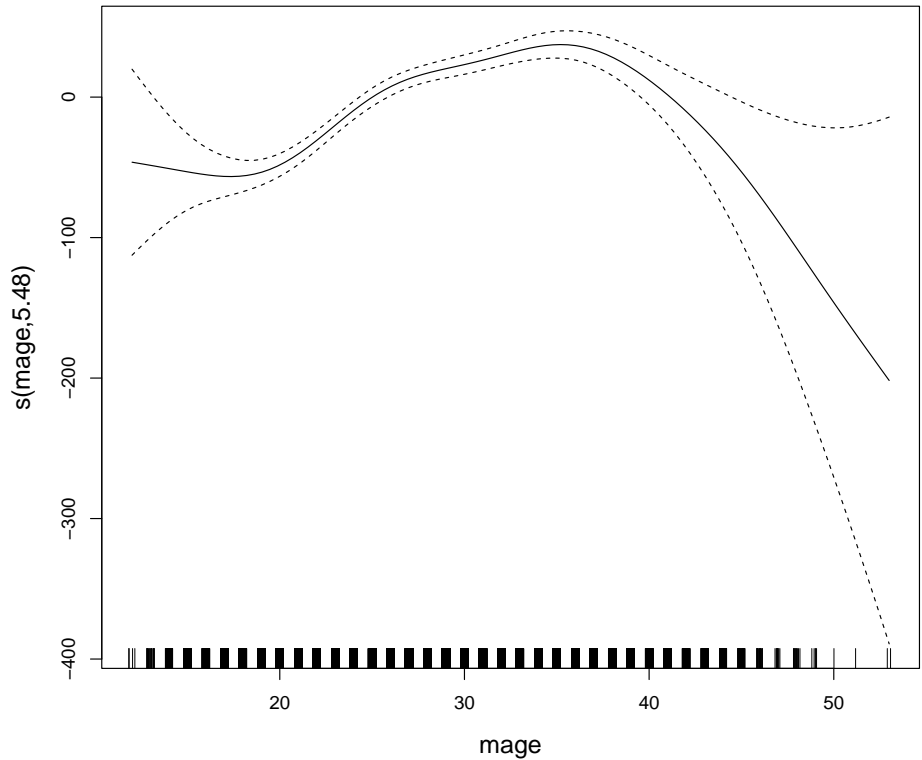

Figure 1: Estimated smooth effect of `mage` on `bwgram` and associated 95% point-wise confidence intervals. The jittered rug plot, at the bottom of each graph, shows the covariate values. The number in brackets in the y-axis caption represents the effective degrees of freedom (edf) of the smooth curve (see Appendix-1 for the definition of edf).

The parametric estimates (obtained using `summary(outC0)`) are consistent with the interpretations that on average an infant of a married mother is 122 grams heavier as compared to an infant of a non-married woman, that an infant born as a multiple birth is on average 894 grams lighter as compared to a single birth, and that on average an infant of a non-Hispanic mother is 55 grams lighter as opposed to that of a Hispanic mother.

Figure 2 (bottom plot) shows the estimated associations (in terms of Kendall's $\hat{\tau}$), averaged by `nonhisp` and `multbirth`, within each county: after accounting for covariates, the association between `bwgram` `wksgest` is present, significant (see last line of code in Appendix-3) and heterogeneous across counties with an average Kendall's $\hat{\tau}$ of $0.25$ and values ranging from $0.20$ to $0.34$.

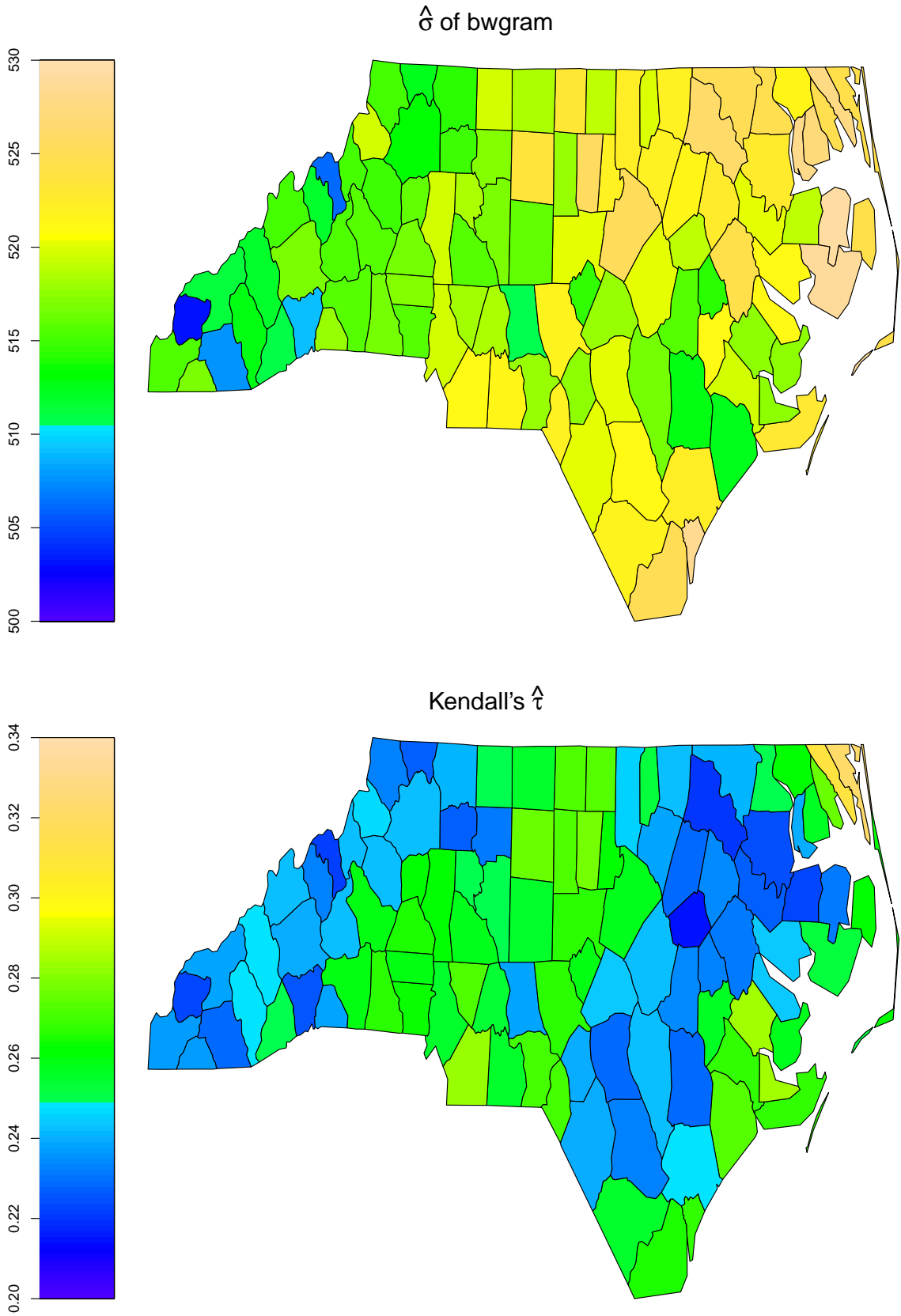

Figure 2: Standard deviations of bwgram and Kendall's $\hat{\tau}$ by county in North Carolina. These have been calculated using the results in Tables 1 and 2 in the main paper after fitting a flexible Clayton copula regression model to birth data.

# 5   R code to simulate data

```
library(copula) # contains archmCopula(), mvdc() and rMvdc()
                # which are needed to simulate from
                # the desired copula
library(gamlss) # contains IG() and GB2() which are
                # needed to employ inverse Gaussian
                # and Singh-Maddala margins
library(SemiParBIVProbit) # contains rMVN() which is useful
                          # for simulating Gaussian correlated
                          # covariates, for instance

cor.cov <- matrix(0.5, 3, 3); diag(cor.cov) <- 1
f1 <- function(x) x * sin(3 * x)
f2 <- function(x) (x + exp(-3*(x-0.5)^2))

data.gen <- function(cor.cov, f1, f2){

cov    <- rMVN(1, rep(0,3), cor.cov)
cov    <- pnorm(cov)

x1 <- cov[,1]
x2 <- cov[,2]
x3 <- round(cov[,3])

eta_mu1     <- 0.5 - 1.25*x2 - 0.8 * x3
eta_mu2     <- 0.1 - 0.9*x1 + f1(x2)
eta_sigma21 <- 1.8
eta_sigma22 <- 0.1
eta_nu      <- 0.2 + x3
eta_theta   <- 0.2 + 0.7*x1 + f2(x2)

Cop <- archmCopula(family = "joe", dim = 2, param = exp(eta_theta) + 1)

speclist1 <- list( mu = exp(eta_mu1), sigma = exp(eta_sigma21) )
speclist2 <- list( mu = exp(eta_mu2), sigma = exp(eta_sigma22), nu = 1,
                   tau = exp(eta_nu) )

spec <- mvdc(copula = Cop, c("IG", "GB2"),
             list(speclist1, speclist2) )

c(rMvdc(1, spec), x1, x2, x3)
}
```

```
dataSim <- matrix(NA, nrow = n, ncol = 5)
for(j in 1:n) dataSim[j,] <- data.gen(cor.cov, f1, f2)
```

Since `archmCopula()` does not allow for the use of vectors for `par`, function `data.gen()` is executed as many times as the number of observations (n) that the user wishes to simulate, as illustrated in the last two lines of code.

# References

Geyer, C. J. (2015). *trust: Trust Region Optimization*. R package version 0.1-6.

Marra, G., Radice, R., Bärnighausen, T., Wood, S. N., & McGovern, M. E. (2016). A simultaneous equation approach to estimating hiv prevalence with non-ignorable missing responses. *Submitted and available upon request*.

Nocedal, J. & Wright, S. J. (2006). *Numerical Optimization*. New York: Springer-Verlag.

Radice, R., Marra, G., & Wojtys, M. (2015). Copula regression spline models for binary outcomes. *Statistics and Computing*, (pp. 1–15).

Wood, S. N. (2004). Stable and efficient multiple smoothing parameter estimation for generalized additive models. *Journal of the American Statistical Association*, 99, 673–686.